\documentclass[fleqn,usenatbib]{mnras}

\usepackage{newtxtext,newtxmath}
\usepackage[T1]{fontenc}

\usepackage{graphicx}
\usepackage{epstopdf}
\usepackage{color}
\usepackage{amsmath}
\usepackage{lastpage}
\usepackage{tikz}

\usepackage[]{hyperref}

\hypersetup{
  colorlinks=true,
  linkcolor=red,
  citecolor=blue,
  bookmarksopen=true,
  bookmarksnumbered=true,
  pdfstartpage={1},  
  pdftitle = {},
  pdfsubject = {},
  pdfauthor = {} ,
  pdfkeywords = {},
  pdfproducer = {LaTeX with hyperref}
}

\newcommand{\kms}{\mathrm{km s}^{-1}}

\def\Bmax{\ifmmode{\>\vert B\vert_{\text{max}}}\else{$\vert B\vert_{\text{max}}$}\fi}
\def\Bnine{\ifmmode{\>\vert B\vert_{90}}\else{$\vert B\vert_{90}$}\fi}

\def\nH{\ifmmode{\>n_{\textnormal{\sc h}}} \else{$n_{\textnormal{\sc h}}$}\fi}

\def\mG{\ifmmode{\>\mu\mathrm{G}}\else{$\mu$G}\fi}
\def\erg{\ifmmode{\> {\rm erg}}\else{erg}\fi}
\def\keV{\ifmmode{\> {\rm keV}}\else{keV}\fi}

\def\deg{\ifmmode{\>^{\circ}}\else{$^{\circ}$}\fi}
\def\onedeg{\ifmmode{\>1^{\circ}}\else{$1^{\circ}$}\fi}

\def\xvir{\ifmmode{\>x_{vir}}\else{$x_{vir}$}\fi}
\def\Mvir{\ifmmode{\>M_{vir}}\else{$M_{vir} $}\fi}
\def\rvir{\ifmmode{\>r_{vir}}\else{$r_{vir}$}\fi}
\def\vvir{\ifmmode{\>v_{vir}}\else{$v_{vir}$}\fi}

\def\tratio{\ifmmode{\>\tau}\else{$\tau$}\fi}

\def\rms{\ifmmode{\>r_{\textnormal{\sc ms}}}\else{$r_{\textnormal{\sc ms}}$}\fi}

\def\Mpc{\ifmmode{\>{\rm Mpc}} \else{Mpc}\fi}
\def\kpc{\ifmmode{\>{\rm kpc}} \else{kpc}\fi}
\def\pc{\ifmmode{\>{\rm pc}} \else{pc}\fi}

\def\Gyr{\ifmmode{\>{\rm Gyr}} \else{Gyr}\fi}
\def\Myr{\ifmmode{\>{\rm Myr}} \else{Myr}\fi}
\def\yr{\ifmmode{\>{\rm yr}} \else{yr}\fi}
\def\pyr{\ifmmode{\>{\rm yr}^{-1}}\else{yr $^{-1}$} \fi}
\def\s{\ifmmode{\>{\rm s}}\else{s}\fi}
\def\ps{\ifmmode{\>{\rm s}^{-1}}\else{s$^{-1}$}\fi}
\def\Hz{\ifmmode{\>{\rm Hz}}\else{Hz}\fi}

\def\kms{\ifmmode{\>{\rm km\,s}^{-1}}\else{km~s$^{-1}$}\fi}

\def\K{\ifmmode{\>{\rm K}}\else{K}\fi}

\def\sr{\ifmmode{\>{\rm sr}}\else{sr}\fi}
\def\psr{\ifmmode{\>{\rm sr}^{-1}}\else{sr$^{-1}$}\fi}
\def\arcs{\ifmmode{\>{\rm arcsec}}\else{arcsec}\fi}
\def\parcs{\ifmmode{\>{\rm arcsec}^{-1}}\else{arcsec${-1}$}\fi}
\def\parcss{\ifmmode{\>{\rm arcsec}^{-2}}\else{arcsec${-2}$}\fi}

\def\cm{\ifmmode{\>{\rm cm}}\else{cm}\fi}
\def\cc{\ifmmode{\>{\rm cm}^{3}}\else{cm$^{3}$}\fi}
\def\sqc{\ifmmode{\>{\rm cm}^{2}}\else{cm$^{2}$}\fi}
\def\pcc{\ifmmode{\>{\rm cm}^{-3}}\else{cm$^{-3}$}\fi}
\def\psc{\ifmmode{\>{\rm cm}^{-2}}\else{cm$^{-2}$}\fi}

\def\g{\ifmmode{\>{\rm g}}\else{g}\fi}
\def\Msun{\ifmmode{\>{\rm M}_{\odot}}\else{M$_{\odot}$}\fi}
\def\hMsun{\ifmmode{\> h^{-1}{\rm M}_{\odot}}\else{$h^{-1}$M$_{\odot}$}\fi}

\def\Zsun{\ifmmode{\>{\rm Z}_{\odot}}\else{Z$_{\odot}$}\fi}

\def\rayl{\ifmmode{\>{\rm R}}\else{R}\fi}
\def\mR{\ifmmode{\>{\rm mR}}\else{mR}\fi}

\renewcommand{\ion}[2]{\hbox{#1\,{\sc #2}}}

\def\lya{\ifmmode{\>{\rm Ly}\alpha}\else{Ly$\alpha$}\fi}

\def\Ha{\ifmmode{\>{\rm H}\alpha}\else{H$\alpha$}\fi}
\def\Hb{\ifmmode{\>{\rm H}\beta}\else{H$\beta$}\fi}

\def\HI{\ifmmode{\> \textnormal{\ion{H}{i}}} \else{\ion{H}{i}}\fi}
\def\HII{\ifmmode{\> \textnormal{\ion{H}{ii}}} \else{\ion{H}{ii}}\fi}
\def\CIV{\ifmmode{\> \textnormal{\ion{C}{iv}}} \else{\ion{C}{iv}}\fi}
\def\SiIV{\ifmmode{\> \textnormal{\ion{S}{iv}}} \else{\ion{Si}{iv}}\fi}

\def\NHI{\ifmmode{\> {\rm N}_{\HI}} \else{N$_{\HI}$}\fi}
\def\MHI{\ifmmode{\> {\rm M}_{ \HI}} \else{M$_{\HI}$}\fi}

\def\mua{\ifmmode{\>\mu_{ \textnormal{\Ha}}}\else{$\mu_{ \textnormal{\Ha}}$}\fi}
\def\alphabha{\ifmmode{\>\alpha_{B}^{(\textnormal{\Ha})}}\else{$\alpha_{B}^{(\textnormal{\Ha})}$}\fi}

\title[HVCs in the magnetised halo]{The role of the halo magnetic field on accretion through High-Velocity Clouds}

\author[A. Gr\o nnow et al.]{
Asger Gr\o nnow,$^{1}$\thanks{E-mail: gronnow@astro.rug.nl}
Thor Tepper-Garc\'{\i}a,$^{2,3,4}$
Joss Bland-Hawthorn$^{2,3}$
and Filippo Fraternali$^{1}$
\\
$^{1}$Kapteyn Astronomical Institute, University of Groningen, 9700AV Groningen, The Netherlands\\
$^{2}$Sydney Institute for Astronomy, School of Physics A28, The University of Sydney, NSW 2006, Australia\\
$^{3}$Centre of Excellence for Astronomy in Three Dimensions (ASTRO-3D), Australia\\
$^{4}$Centre for Integrated Sustainability Analysis, School of Physics A28, The University of Sydney, NSW 2006, Australia\
}

\date{Accepted 2021 Nov 23. Received 2021 Nov 1; in original form 2021 Aug 30}

\pubyear{2021}

\begin{document}
\label{firstpage}
\pagerange{\pageref{firstpage}--\pageref{lastpage}}
\maketitle

\begin{abstract}
High-Velocity Clouds (HVCs) are believed to be an important source of gas accretion for star formation in the Milky Way. Earlier numerical studies have found that the Galactic magnetic field and radiative cooling strongly affects accretion. However, these effects have not previously been included together in the context of clouds falling through the Milky Way's gravitational potential. We explore this by simulating an initially stationary cloud falling through the hot hydrostatic corona towards the disc. This represents a HVC that has condensed out of the corona. We include the magnetic field in the corona to examine its effect on accretion of the HVC and its associated cold gas. Remnants of the original cloud survive in all cases, although a strong magnetic field causes it to split into several fragments. We find that mixing of cold and hot gas leads to cooling of coronal gas and an overall growth with time in cold gas mass, despite the low metallicity of the cloud and corona. The role of the magnetic field is to (moderately to severely) suppress the mixing and subsequent cooling, which in turn leads to less accretion compared to when the field is absent. A stronger field leads to less suppression of condensation because it enhances Rayleigh-Taylor instability. However, magnetic tension in a stronger field substantially decelerates condensed cloudlets. These have velocities typically a factor 3--8 below the velocity of the main cloud remnants by the end of the simulation. Some of these cloudlets likely disperse before reaching the disc.
\end{abstract}




\section{Introduction}
\label{sec:intro}
Inflow of cold gas is needed in the Milky Way (MW) to explain the observed stellar chemical abundances \citep{vandenbergh62,schlesinger12}, the chemical evolution of the ISM \citep{larson72,edmunds90,schoenrich17}, and the current star formation rate of $\sim 1 M_\odot$ yr$^{-1}$ \citep{robitaille10,licquia15} and star formation history. In general, this gas is not observed directly, neither in the MW nor in other star forming galaxies, presumably due to it being largely diffuse \citep[but see][]{zheng17,koch18}. However, infalling High-Velocity Clouds (HVCs) in the MW do offer a clear indication of gas accretion. These are structures of cool atomic hydrogen with velocities that significantly differ from the local standard of rest (LSR, typically defined as $|v_{\mathrm{LSR}}| > 90 \kms$) first observed by \cite{muller63}. They are classically associated with H\textsc{i}, i.e. neutral hydrogen, although partially and fully ionised HVCs are also observed \citep{sembach03,lehner12,richter15}. Although estimates of the total mass in infalling HVCs typically fall short of what would be needed to fully replenish the gas lost to star formation in the MW disc, they are still estimated to represent a significant fraction of gas accretion \citep{putman12,fox19}.

HVCs are generally found to be within MW's hot diffuse gas corona based on distance constraints of typically $1 \lesssim d \lesssim 50$ kpc \citep[e.g.][]{wakker01,putman03b,thom08,lehner10,peek16}, less than the expected extent of the corona \citep{faerman17,bregman18}, and signs of interaction with an external medium. These are seen in the head-tail morphology of many HVCs \citep{putman11}. Recently, indications of hydrodynamical instabilities associated with cloud-corona interaction have also been found in detailed morphological analysis of Complex A \citep{barger20}. The origins of HVCs are still not clear, but in any case they appear to have multiple formation mechanisms. The metallicities generally observed in HVCs are too low for them to originate in the disc.  The exception to this is the Smith Cloud \citep{smith63,fox16} that may either represent outflowing entrained gas \citep{marasco17}; a small dark matter (DM) halo that has collided with the disc and accreted gas from the ISM \citep{nichols09,nichols14}, or a `streamer' in the form of a gaseous structure dislodged from the Galactic gas disc by a transiting, gas-bearing DM mini-halo \citep{tepper-garcia18a,galyardt16}.

Some HVCs, such as the ones in the Magellanic Stream, appear to have been stripped from satellite galaxies. However, the infall of the Sagitarius Dwarf galaxy does not seem to have led to the formation of any currently observed HVCs \citep{tepper-garcia18b}. A plausible origin for other HVCs is that they form through condensation, i.e. cooling of coronal gas.
This can be triggered by outflows, as claimed by \cite{fraternali15} as a possible origin of Complex C at $z=8.4$ kpc above the disc, or by density perturbations in the corona through thermal instability typically around $z\sim 10$ kpc \citep{maller04,joung12,sormani19}.

Buoyant oscillations can disrupt this process \citep{binney09}, however the Galactic halo magnetic field has been shown to stabilise perturbations against this \citep{ji18}. The halo magnetic field also affects the interaction between the corona and HVCs as they are traveling through it. This magnetic field is difficult to constrain but has a significant ordered component that appears to mainly curl around the $z$-axis (the axis pointing perpendicularly away from the disc) with strength decreasing with $z$ \citep{sunreich10,jansson12a,unger19}.

The general scenario of a cloud in relative motion with a surrounding hot magnetised medium has been widely studied in numerical simulations in the literature \cite[e.g.][]{jones96,gregori99,gregori00,santillan99,dursi08,kwak09,mccourt15,banda-barragan16,gronnow17,gronnow18,banda-barragan18,cottle20,sparre20}. In most of these simulations \cite[the exceptions being ][]{santillan99,kwak09} there is no external gravitational potential and the cloud is instead given an initial velocity, either directly or, more commonly, by injecting a constant velocity `wind' around it. We refer to those as `wind tunnel' simulations. They are usually analysed in the context of entrained gas in outflows, but can equivalently be interpreted as inflowing gas. However, while such simulations can provide useful insights for infalling gas, the velocity evolution is quite different from that of an HVC condensed out of the corona. With no gravitational potential present, the cloud decelerates due to drag from the outset, eventually coming to rest with respect to the surrounding gas (in the wind frame this corresponds to the cloud \emph{accelerating} until it becomes comoving with the wind). 

The initial high relative velocity leads to an initial shock propagating through the cloud, which is appropriate for entrained gas but not generally for infalling clouds \citep{bland-hawthorn07,tepper-garcia15}. Typically this shock, in combination with Kelvin-Helmholtz (KH) and Rayleigh-Taylor (RT) instabilities, destroys the cloud before it becomes comoving with the surrounding medium. The KH and RT instabilities are caused by the velocity shear between the cloud and the surrounding gas and parcels of gas at different densities being accelerated into each other, respectively. Cloud destruction in outflows is a big topic in contemporary astrophysics because entrained clouds are generally shredded before they can accelerate and travel sufficiently to agree with observational constraints \citep{zhang17}. 

Overall, the cloud's life time roughly scales with the density contrast between the cloud and surrounding medium $\chi$, relative velocity $v$, and size of the cloud $L$ as $\sim \chi^{1/2}v L^{-1}$ \citep{jones96} while the dependence on the magnetic field is more complicated. The first three-dimensional simulation of this kind, \cite{gregori99}, found that the magnetic field in fact \emph{hastened} the destruction by enhancing RT instability along the third dimension, i.e. the direction perpendicular to both the wind and the magnetic field. In contrast, later higher resolution simulations run with a variety of codes and numerical schemes
\citep[e.g.][]{mccourt15,banda-barragan16,gronnow17,banda-barragan18,gronnow18,cottle20,gronke20a,li20} have found that the magnetic field generally does not hasten the destruction. Instead, it tends to extend the overall cloud life time through its partial suppression of KH instability, although the effect is limited and often by itself insufficient to solve the problem of shorter than expected survival.

However, in the context of accretion of infalling gas, the survival of the remnant of the original cloud is not the most relevant diagnostic. Rather, the total mass of cold gas associated with the cloud, whether in the form of a single cloud or a complex containing many cloudlets, is the main quantity of interest. In this case, cloud stripping can even lead to an overall increase in cold gas mass through condensation as shown in simulations that include radiative cooling \citep[e.g.][]{marinacci10,armillotta16,gronke20a}. The mixing of stripped cold gas with the hot coronal gas can, in some realistic parts of the parameter space of cloud and corona density, velocity, temperature, and metallicity, lead to efficient condensation that increases the total cold gas mass, rather than evaporation of the stripped gas. This process is typically associated with intermediate-velocity,
relatively metal-enriched galactic fountain clouds travelling through the inner part of the corona. In this context, \cite{gronnow18} showed that the halo magnetic field strongly restricts the efficiency of condensation while still allowing growth in cold gas mass. However, \cite{gritton17} claimed that even essentially pristine HVCs can lead to efficient condensation based on wind tunnel simulations without magnetic fields. More generally, \cite{gronke20a} and \cite{li20} have recently investigated the parameter space that leads to growth and destruction of cold gas in magnetized wind tunnel simulations and find that growth occurs for a large part of the parameter space relevant to HVCs.

In this work, we investigate the evolution of cold gas in an HVC falling through the MW's magnetised hot corona using magnetohydrodynamic (MHD) simulations. We simulate an initially stationary cold cloud surrounded by a hydrostatic hot corona with density and magnetic field strength increasing towards the disc and a uniform gravitational acceleration. Due to the external potential the cloud accelerates and eventually reaches HVC velocities. This is representative of an HVC formed through thermal instability in the corona.
This setup is reminiscent of the one used by \cite{heitsch09} {\it but with the important addition of the halo magnetic field.} Clouds falling through a hydrostatic medium with a magnetic field have been simulated before by \cite{santillan99} and \cite{kwak09}. However, both of those studies focused on clouds closer to the disc and, crucially, neglected radiative cooling and thus were unable to follow the cold gas evolution.

This paper is organised as follows: In Section \ref{sect:ICs} we describe our numerical setup and methods. In Section \ref{sec:results} we show the results of these simulations which we discuss in the context of cold gas evolution in Section \ref{sec:discussion}. Finally, we conclude in Section \ref{sec:summary}. 


\section{Numerical setup}
\label{sect:ICs}

\subsection{Initial conditions}
We simulate a cold ($T = 10^4$ K) cloud initially at rest with the hot corona and follow its evolution as it accelerates and becomes an HVC\footnote{We assume that the velocity w.r.t. the corona is representative of the LSR velocity such that the cloud can be referred to as an HVC in our simulations once it reaches $|v_z| > 90 \kms$.} due to the gravitational potential of the MW disc. Our simulations are in a three-dimensional cartesian coordinate system $(x,y,z)$ where the $z$ axis is perpendicular to the disc. All gas is assumed to be monatomic ideal gas and so has an adiabatic index of $\gamma=5/3$. We model the hot coronal gas along the path of the cloud as being in (magneto)hydrostatic equilibrium. When there is no magnetic field we assume an isothermal confining medium with temperature $T_\mathrm{iso}=2 \times 10^6$ K typical of the Galactic corona \citep{henley15,miller15}. We assume a constant gravitational acceleration of the Milky Way disc of $g=-10^{-8}$ cm s$^{-2}$ in the $z$ (height above the disc) direction. This is appropriate for our range of 2 kpc $\leq z \leq$ 10 kpc in the solar neighbourhood according to the models of
\cite{kalberla08,joung12}. We have additionally confirmed that this value is consistent with the `MWPotential2014' model of the Galactic potential of \cite{bovy15} for virial masses in the range $1.0-1.4 \times 10^{12} M_\odot$ as indicated by observational constraints \citep[e.g.][]{posti19,cautun20}. For the solar cylindrical radius assumed in this model of $R=8$ kpc the gravitational acceleration is always within 20 per cent of our assumed value for $2 \leq z \leq 10$ kpc. Consequently, the density is only a function of $z$ and the resulting particle density profile becomes
\begin{equation}
\label{eqn:densprof}
n_h(z) = n_{h,0}\exp{\left(\frac{-g \mu z}{k_B T_\mathrm{iso}}\right)},
\end{equation}
where $k_B$ is the Boltzmann constant and $\mu$ is the mean molecular weight. We use the subscript `h' for `halo' to refer to what we otherwise call the corona to avoid confusion with `c' which we use to refer to the cloud. We choose $n_{h,0}=0.01 \pcc$ to obtain densities that agree with the loose constraints on the coronal density at 2 kpc $\leq z \leq 10$ kpc in the literature \citep{stanimirovic02,bregman07,grcevich09,miller15}.

The magnetic field of the halo is poorly known but appears to have a significant, mostly axisymmetric, ordered component that falls off with $z$ and cylindrical radius $R$ \citep{sunreich10,jansson12a}. We base our magnetic field on the model of \cite{sunreich10} in the limit of negligible differences in $R$. That is, we assume that the field does not change significantly in magnitude or direction along the sub-kpc extent along $R$ of the cloud and so is only a function of $z$. We align the field such that it points in the $+x$ direction. This choice is inconsequential due to the $xy$-symmetry in the initial conditions. This field is thus
\begin{align}
B(z)	= & \frac{B_0}{1+\left[(\vert z \vert-z_a) / z_b \right]^2},
\end{align}
with $z_a=1.5$ kpc and $z_b=4$ kpc. We use a `weak' and a `strong' field setup with $B_{0, \mathrm{ weak}}= 0.3 \mu$G and $B_{0, \mathrm{ strong}}=5B_{0, \mathrm{ weak}} = 1.5 \mu$G, respectively. Because this field initially has no tension force due to all field lines being straight, it simply acts as another pressure term with magnetic pressure $P_\mathrm{mag}(z)=B(z)^2/8\pi$. It is, however, not a force-free field due to its gradient in $z$ and so our isothermal non-magnetic setup must be modified when the field is added. We choose to retain the density profile of Eq. \ref{eqn:densprof}, instead allowing the temperature to vary with $z$ to compensate for the added force in the $+z$ direction from the magnetic pressure gradient. Hence, the gas pressure profile required for equilibrium is
\begin{equation}
P(z) = n_h(z)k_B T_\mathrm{iso} - P_\mathrm{mag}(z)
\end{equation}
and the temperature becomes
\begin{equation}
T_h(z) = T_\mathrm{iso} - P_\mathrm{mag}(z)/(n_h(z)k_B).
\end{equation}
However, due to the gas pressure dominating, the magnetised corona is still nearly isothermal with $0.93 T_{\mathrm{iso}} < T_h(z) \leq T_{\mathrm{iso}}$ for all $z$.

We initialise the cloud as a spherical overdensity centred at a height of $z_0=10$ kpc above the disc in pressure equilibrium with the hot corona. The density profile as function of radius from the cloud centre $r$ is a smoothed top-hat
\begin{equation}
\label{eq:densprofile}
	n(r,z)=n_h(z) + \frac{1}{2}(n_c - n_h(z))\left\{1 - \tanh{\left[ s \left(\frac{r}{r_c}-1\right)\right]}\right\} \, ,
\end{equation}
where $n$ is total particle density, $n_c$ is the central cloud particle density, $r_c$ is the cloud radius, and $s=10$ sets the steepness of the profile. Hence, the cloud radius is defined as the radius where the density $n(r_c)$ is halfway between $n_c$ and $n_h$ which is $n(r_c) \approx n_c/2$ because $n_c \gg n_h$. The $z$ dependence is to a very good approximation only a dependence on the height at the cloud centre $z_0$ because the $\Delta z=2r_c$ range covered by the cloud is much smaller than the scale height of the corona. Hence, the density profile of the cloud is approximately spherically symmetric. The initial mass of the cloud is about $7.6\times 10^4$ $M_\odot$. We assume that the cloud and the corona both have a uniform metallicity of $0.2 Z_{\odot}$ in agreement with observations of HVCs and the MW corona \citep{shull09,miller15,hodges-kluck18}. We add a tracer quantity to the cloud that is passively advected with the flow. That is, the tracer is set to 1 within $r_c$ and 0 elsewhere. This allows us to separate cold gas that was originally cold cloud material from cold gas that has condensed from hot gas during in the simulation. We always set the initial temperature of the cloud $T_c$ to equal the cooling floor at $10^4$ K to ensure its stability (see Section \ref{sec:nummethod}). Due to the pressure equilibrium, $n_c$ then follows from $n_h$ and $T_h$. We summarise the physical parameters that are the same in all simulations in Table \ref{tab:params}.

\begin{table}
	\centering
	\caption{Physical parameters that are the same in all of our simulations.}
	\label{tab:params}
	\begin{tabular}{ccccccc}
        $n_{h,0}^a$ & $g$ & $T_{\mathrm{iso}}$ & $T_c$ & $Z/Z_{\odot}$ & $r_c$ & $\chi\equiv\rho_c/\rho_h^b$\\
        (cm$^{-3}$) & (cm s$^{-2}$) & (K) & (K) & & (kpc) &\\
        \hline
        $10^{-2}$ & $10^{-8}$ & $2 \times 10^6$ & $10^4$ & 0.2 & 0.1 & 412\\
        \hline
        \multicolumn{7}{l}{\footnotesize$^a$ The halo particle density at the cloud's initial height is}\\
        \multicolumn{7}{l}{\footnotesize $n_h(10$ kpc$) = 3.4 \times 10^{-3}$ cm$^{-3}$. For the given density contrast this}\\
        \multicolumn{7}{l}{\footnotesize leads to a cloud particle density of $n_c = 0.682$ cm$^{-3}$.}\\
        \multicolumn{7}{l}{\footnotesize This varies slightly across the $\Delta z=0.2$ kpc size of the cloud.}\\
        \multicolumn{7}{l}{\footnotesize$^b$ The \emph{particle} density contrast $n_c/n_h\approx 200$ is given by $T_h/T_c$ due to}\\
        \multicolumn{7}{l}{\footnotesize pressure equilibrium and is approximately half that of $\chi$ due to}\\
        \multicolumn{7}{l}{\footnotesize differences in the mean molecular weight. This value is approximately}\\
        \multicolumn{7}{l}{\footnotesize equal in the HD and MHD simulations since $T_h(10 \kpc)\approx T_{\mathrm{iso}}$.}\\
        \hline
    \end{tabular}
\end{table}

\begin{table}
	\centering
	\caption{The magnetic field strength normalisation, resolution, final time, and final $z$ coordinate of the leading edge of the main remnant in the simulation runs. Note that the final times and heights in the simulations are constrained by our corona model and numerical effects as described in Section \ref{sec:nummethod}. They are hence not related to the state of the clouds which in all cases still exist at the end of the simulations. Each simulation has a short acronym that is used when referring to them throughout the paper.}
	\label{tab:sims}
	\begin{tabular}{lcccc}
        \hline
        Name & $B_0$ & Resolution & $t_{\mathrm{end}}$ & $z_c(t_{\mathrm{end}})$\\
         & ($\mu$G) & (cells/$r_c$) & (Myr) & (kpc)\\
        \hline
        HD & 0 & 50 & 75 & 2\\
        MHD-W & 0.3$^a$ & 50 & 68 & 3.5\\
        MHD-S & 1.5$^b$ & 50 & 64 & 4\\
        MHD-Sl & 1.5 & 25 & 59 & 5\\
        MHD-Sh & 1.5 & 100 & 43 & 7\\
        \hline
        \multicolumn{5}{l}{\footnotesize$^a$ In terms of $\beta\equiv P/P_{\mathrm{mag}}$ this corresponds to $\beta\approx 7700$ at $z=10$ kpc}\\
        \multicolumn{5}{l}{and $\beta \approx 640$ at $z=2$ kpc.}\\
        \multicolumn{5}{l}{\footnotesize$^b$ In terms of $\beta\equiv P/P_{\mathrm{mag}}$ this corresponds to $\beta\approx 310$ at $z=10$ kpc}\\
        \multicolumn{5}{l}{and $\beta \approx 25$ at $z=2$ kpc.}\\
        \hline
    \end{tabular}
\end{table}

\subsection{Numerical methods}
\label{sec:nummethod}
We use the RAMSES Adaptive Mesh Refinement (AMR) code \citep{teyssier02,fromang06} to evolve our simulations in a cartesian domain of $8\times 8\times 8$ kpc. We assume ideal MHD and evolve the magnetic field with a constrained transport scheme which guarantees that the field is divergence free to machine precision. The use of AMR allows us to use a domain sufficiently large to follow the entire tail of stripped material from the cloud throughout the simulations at high resolution. There is a factor of 2 difference in the maximum number of cells along each dimension between the AMR levels, with the coarsest level having $64^3$ cells and the finest level having up to $4096^3$ cells for our standard resolution runs. Thus the maximum resolution is $\Delta x\approx 2$ pc or about one fiftieth of the initial radius of the cloud. We refine based on the density gradient to also capture the mixing of relatively low density gas in the wake. Our simulation domain is much wider than necessary along the transverse ($x$ and $y$) directions because RAMSES does not allow for simulation domains with unequal lengths. However, due to our use of AMR the corona far from the cloud is always on the coarsest level and so the additional computational resources used in evolving this large domain is negligible. We only include the relevant part of the volume in our slice and projection plots in Section \ref{sec:results}.

Initially, the simulation box covers 8 kpc $< z < 16$ kpc. We keep the simulation approximately in the HVC's rest frame by regularly subtracting the centre of mass velocity of the cloud material. We keep track of the cloud material through the use of a passive scalar (i.e. a tracer fluid quantity) initially set to 1 for $r<r_c$ and 0 elsewhere, which we denote as $C$. This comoving cloud frame method has the advantage that the inner parts of the corona, where the density and magnetic field strengths are relatively high and the MHD equilibrium is easily disturbed by numerical effects, is only included near the end of the simulations. It also reduces truncation errors related to advection by minimising the overall velocity of the cloud material with respect to the computational mesh. We discuss this method and compare the evolution in this frame to that of the corona reference frame in Appendix \ref{sec:refframe}. We include optically thin radiative cooling assuming collisional ionisation equilibrium down to a temperature floor of $10^4$ K using the default RAMSES cooling tables calculated using CLOUDY \citep{ferland98}. This takes the temperature dependence of $\mu$ into account. In order to keep the corona stable we disable cooling in unmixed coronal gas. This is justified because the unmixed coronal gas, due to its low density and high temperature (that is also near the minimum in the cooling curve) has a much greater cooling time scale than that of the mixed gas and typically longer than the time span of the simulations. For this we define a passive scalar initially set to 1 where $\rho>2\rho_h$ and 0 elsewhere and only allow cooling in cells where this tracer is greater than zero. This does not affect coronal gas condensation because this process is driven by mixing with cloud material.

In Table \ref{tab:sims} we show the different magnetic field strengths and resolutions for our five different simulation runs. We also note the time and approximate $z$ coordinate of the front of the HVC at the end of each simulation. For the simulation without a magnetic field, HD, we stop the simulation once the main cloud reaches the disc-corona interface at roughly $z\approx 2$ kpc \citep{haffner03,gaensler08}. As can be seen, we are not able to follow the HVC all the way to this point in simulations that include the halo magnetic field. This is due to the strongly amplified draped field at the front of the main cloud remnant. The combination of a very strong magnetic field and a low density (as this region is in the corona just in front of the cloud) leads to a very high velocity of fast magnetosonic waves. This in turn leads to very short simulation time steps which results in the simulations becoming unfeasibly expensive to run past a certain time. However, the two simulations with a magnetic field at our standard resolution, MHD-W and MHD-S, still ran for 90 and 85 per cent of the time and 81 and 63 per cent of the distance, respectively, of simulation HD. Hence, we expect the monotonic increases that we see in all cases for the cold and low velocity gas mass described in Section \ref{sec:results} to be qualitatively representative of the full evolution.

\section{Results}
\label{sec:results}

\begin{figure*}
    \centering
    \vspace{4cm}
    \begin{picture}(75,100)
    \put(0,0){\includegraphics[height=0.4\textwidth]{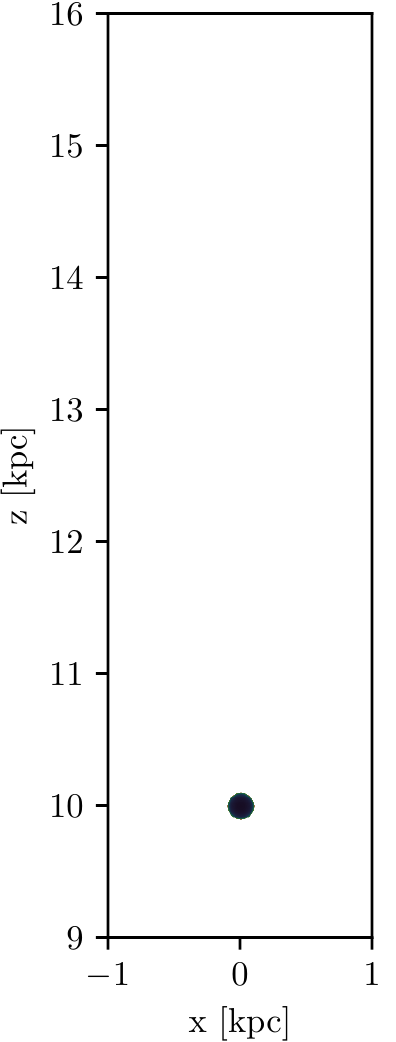}}
    \put(24,190){HD}
    \put(24,182){$t=0$ Myr}
    \end{picture}
    \begin{picture}(68,100)
    \put(0,0){\includegraphics[height=0.4\textwidth]{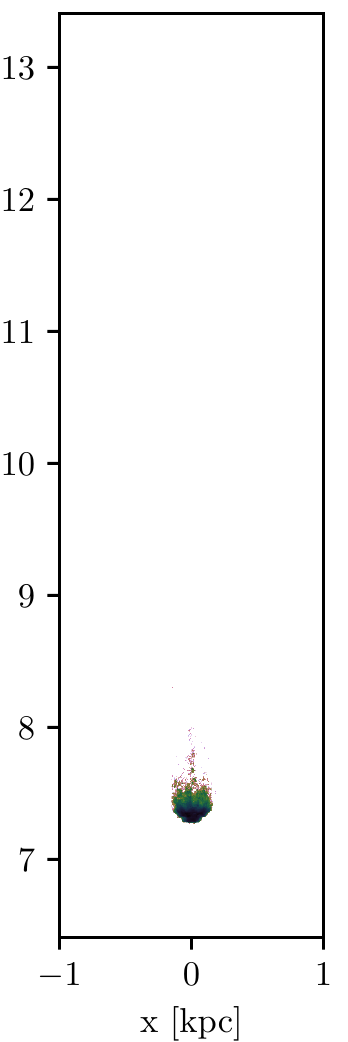}}
    \put(15,190){HD}
    \put(15,182){$t=41$ Myr}
    \end{picture}
    \begin{picture}(68,100)
    \put(0,0){\includegraphics[height=0.4\textwidth]{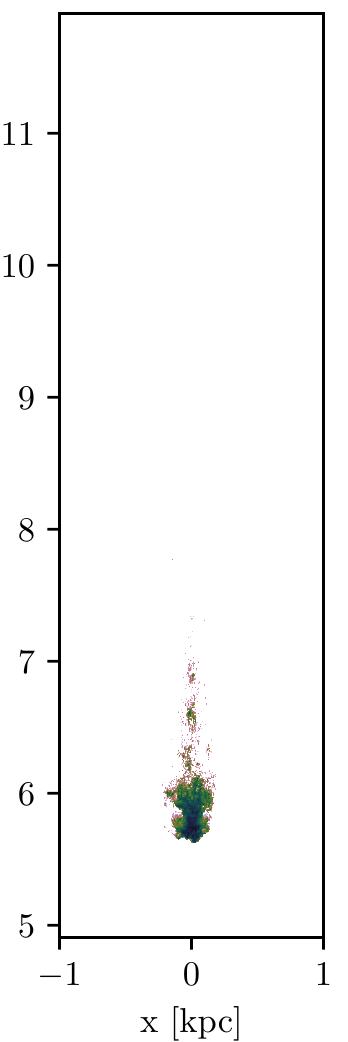}}
    \put(15,190){HD}
    \put(15,182){$t=52$ Myr}
    \end{picture}
    \begin{picture}(68,100)
    \put(0,0){\includegraphics[height=0.4\textwidth]{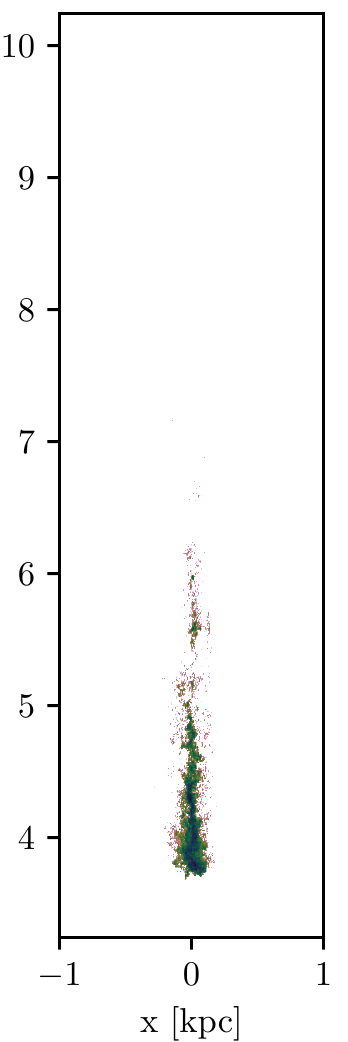}}
    \put(15,190){HD}
    \put(15,182){$t=64$ Myr}
    \end{picture}
    \begin{picture}(75,100)
    \put(0,0){\includegraphics[height=0.4\textwidth]{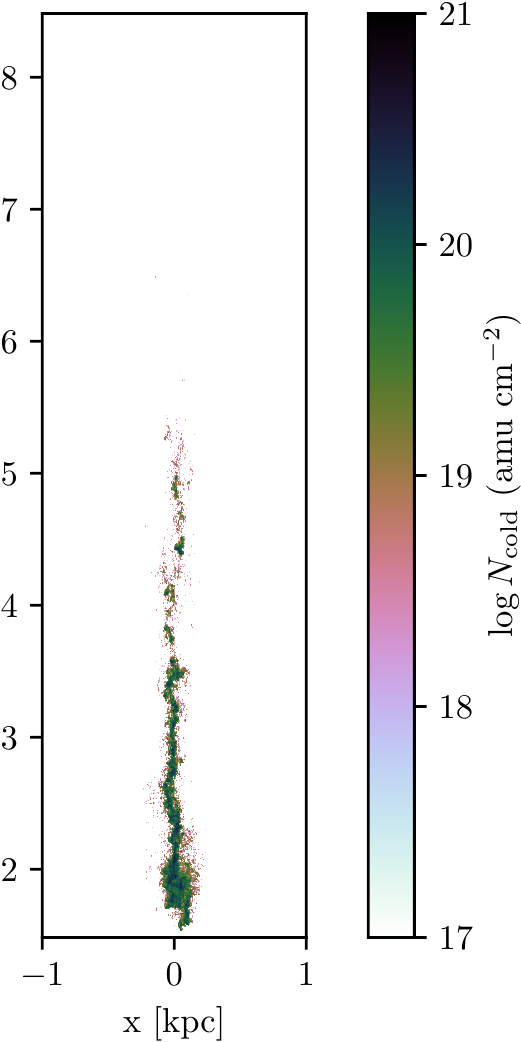}}
    \put(15,190){HD}
    \put(15,182){$t=75$ Myr}
    \end{picture}\\
    \vspace{4cm}
    \begin{picture}(75,100)
    \put(0,0){\includegraphics[height=0.4\textwidth]{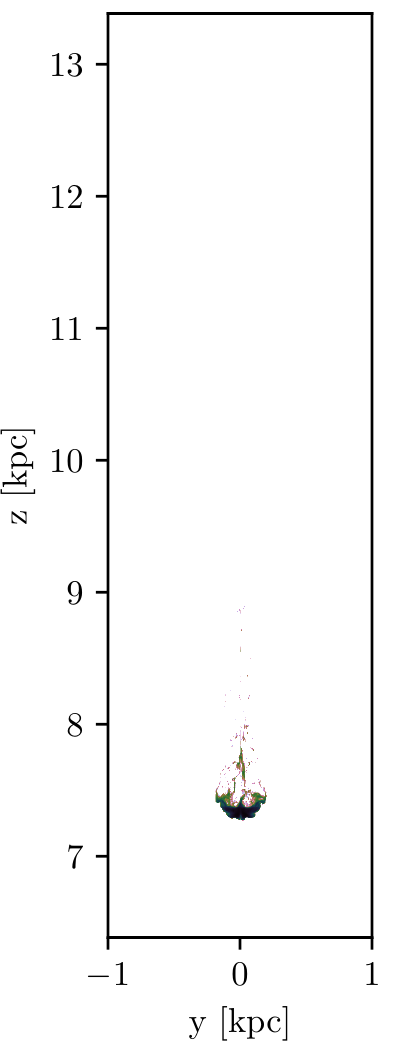}}
    \put(24,190){MHD-W}
    \put(24,181){$x$ projection}
    \put(24,172){$t=41$ Myr}
    \end{picture}
    \begin{picture}(65,100)
    \put(0,0){\includegraphics[height=0.4\textwidth]{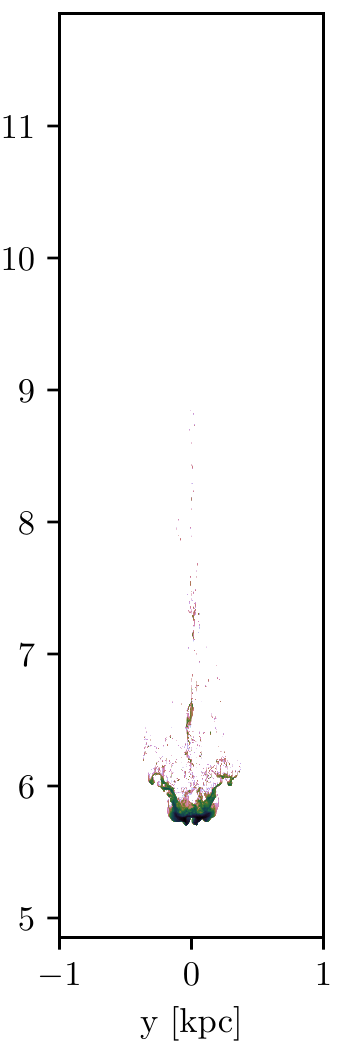}}
    \put(15,190){MHD-W}
    \put(15,181){$x$ projection}
    \put(15,172){$t=52$ Myr}
    \end{picture}
    \begin{picture}(70,100)
    \put(0,0){\includegraphics[height=0.4\textwidth]{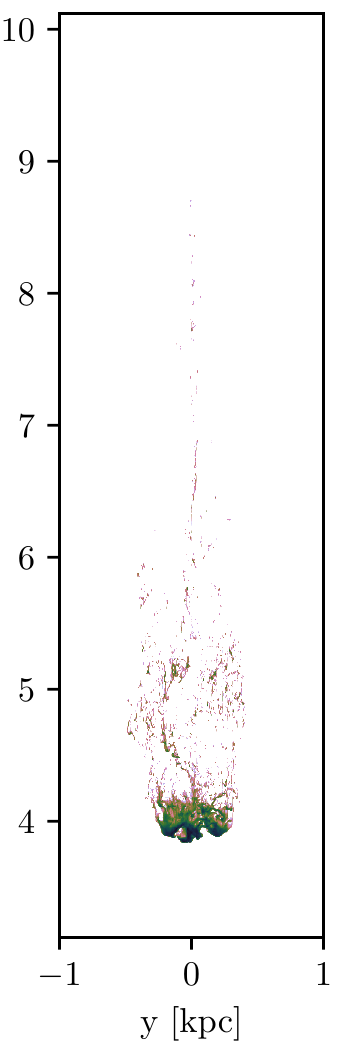}}
    \put(15,190){MHD-W}
    \put(15,181){$x$ projection}
    \put(15,172){$t=64$ Myr}
    \end{picture}
    \hspace{1cm}
    \begin{picture}(75,100)
    \put(0,0){\includegraphics[height=0.4\textwidth]{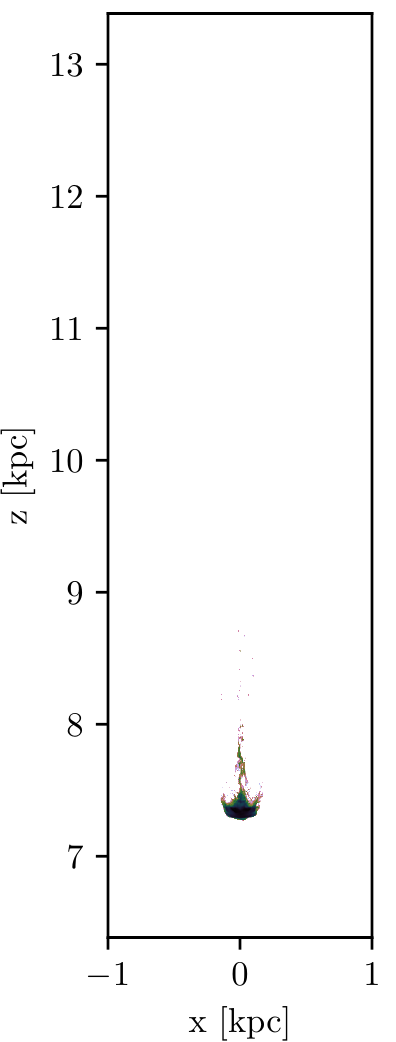}}
    \put(24,190){MHD-W}
    \put(24,181){$y$ projection}
    \put(24,172){$t=41$ Myr}
    \end{picture}
    \begin{picture}(65,100)
    \put(0,0){\includegraphics[height=0.4\textwidth]{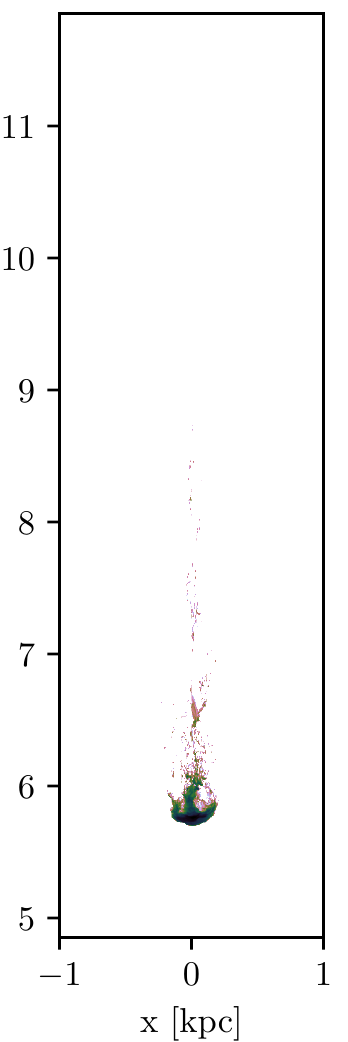}}
    \put(15,190){MHD-W}
    \put(15,181){$y$ projection}
    \put(15,172){$t=52$ Myr}
    \end{picture}
    \begin{picture}(70,100)
    \put(0,0){\includegraphics[height=0.4\textwidth]{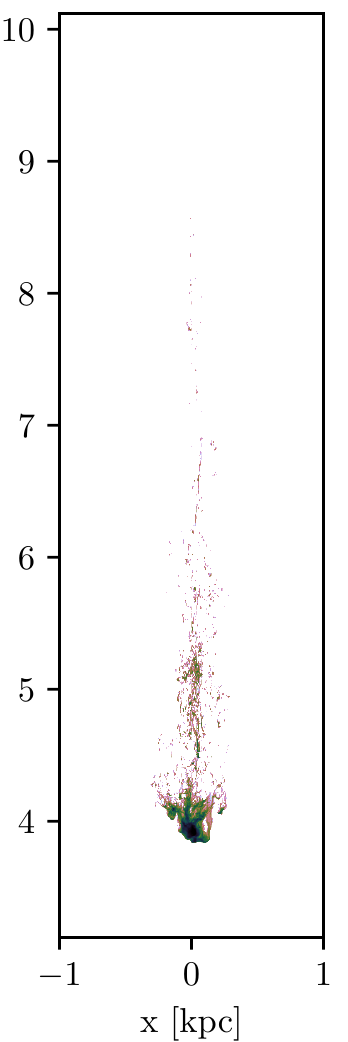}}
    \put(15,190){MHD-W}
    \put(15,181){$y$ projection}
    \put(15,172){$t=64$ Myr}
    \end{picture}\\
    \vspace{4cm}
    \begin{picture}(75,100)
    \put(0,0){\includegraphics[height=0.4\textwidth]{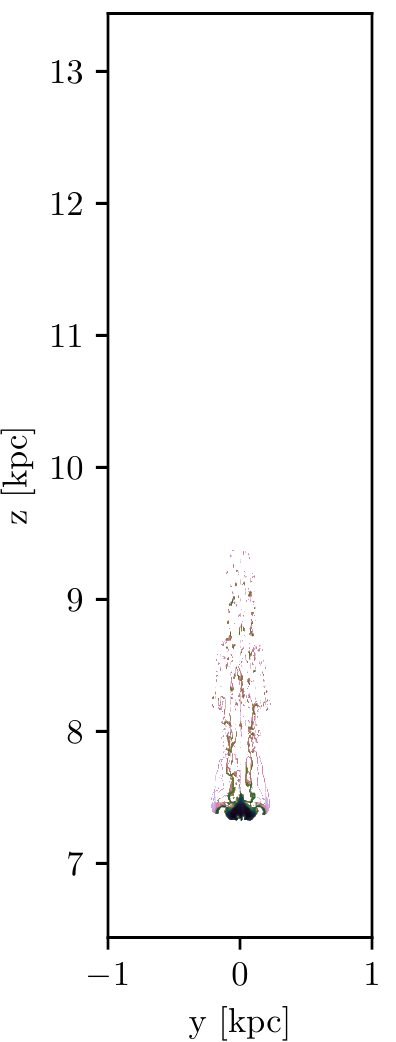}}
    \put(24,190){MHD-S}
    \put(24,181){$x$ projection}
    \put(24,172){$t=41$ Myr}
    \end{picture}
    \begin{picture}(65,100)
    \put(0,0){\includegraphics[height=0.4\textwidth]{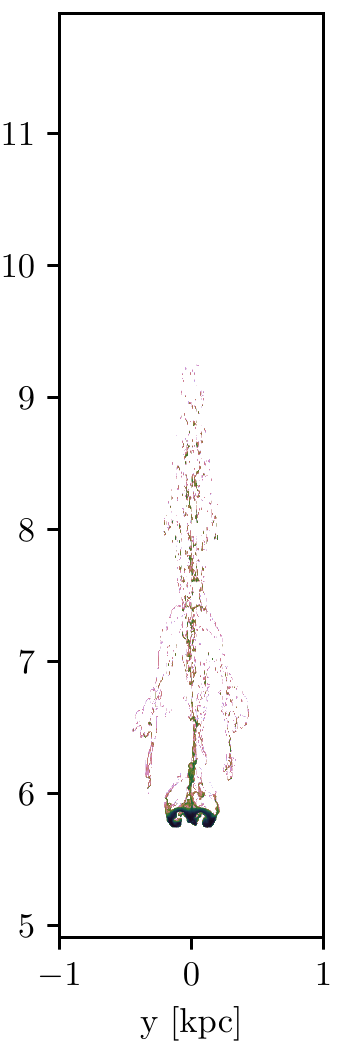}}
    \put(15,190){MHD-S}
    \put(15,181){$x$ projection}
    \put(15,172){$t=52$ Myr}
    \end{picture}
    \begin{picture}(70,100)
    \put(0,0){\includegraphics[height=0.4\textwidth]{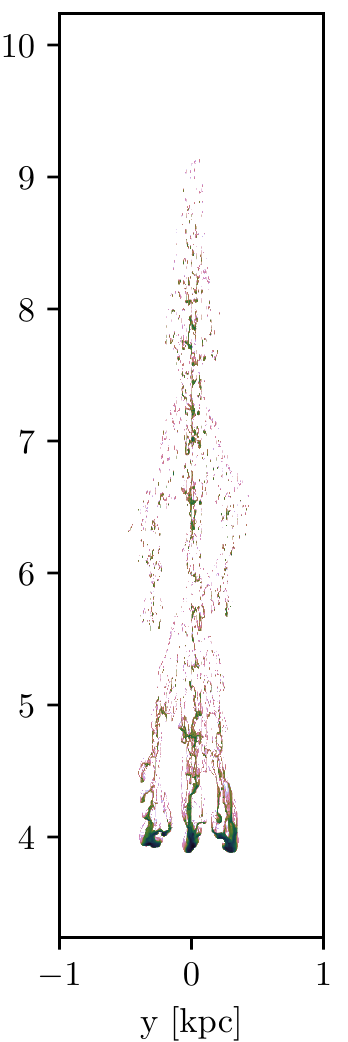}}
    \put(15,190){MHD-S}
    \put(15,181){$x$ projection}
    \put(15,172){$t=64$ Myr}
    \end{picture}
    \hspace{1cm}
    \begin{picture}(75,100)
    \put(0,0){\includegraphics[height=0.4\textwidth]{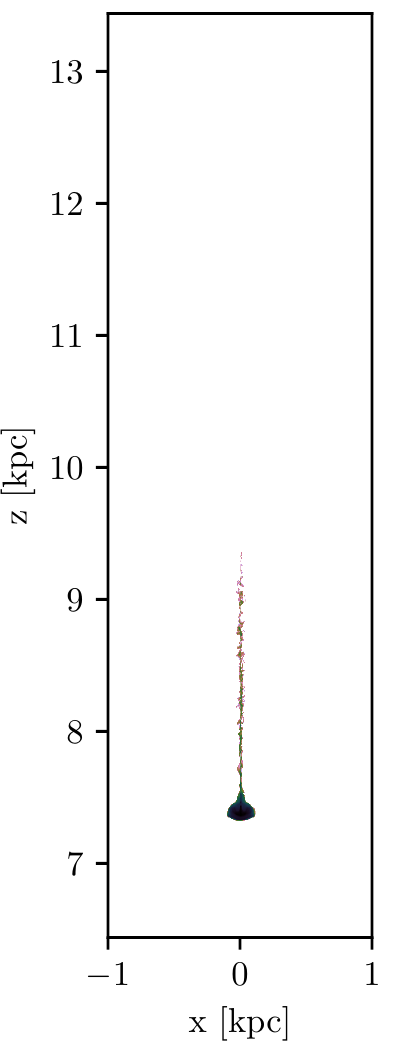}}
    \put(24,190){MHD-S}
    \put(24,181){$y$ projection}
    \put(24,172){$t=41$ Myr}
    \end{picture}
    \begin{picture}(65,100)
    \put(0,0){\includegraphics[height=0.4\textwidth]{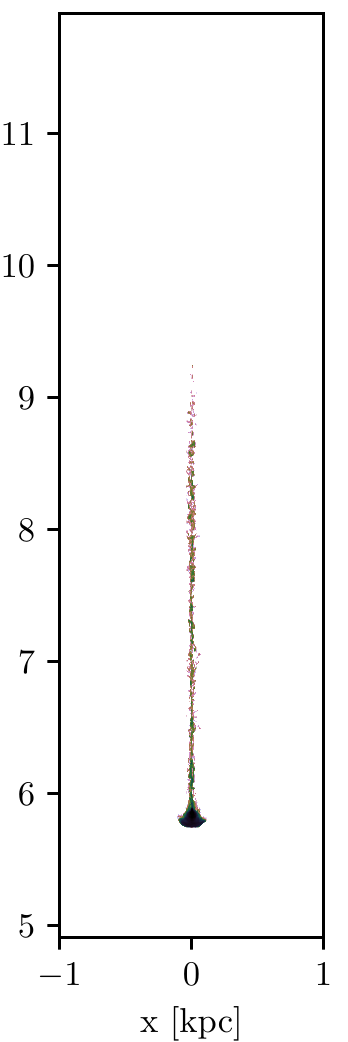}}
    \put(15,190){MHD-S}
    \put(15,181){$y$ projection}
    \put(15,172){$t=52$ Myr}
    \end{picture}
    \begin{picture}(70,100)
    \put(0,0){\includegraphics[height=0.4\textwidth]{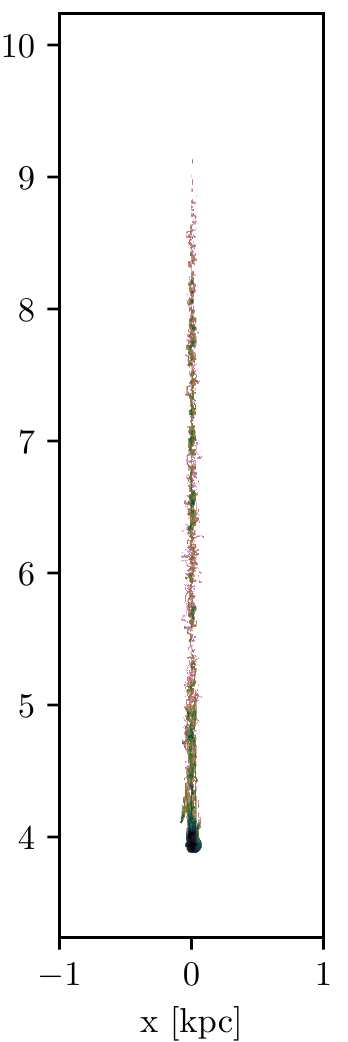}}
    \put(15,190){MHD-S}
    \put(15,181){$y$ projection}
    \put(15,172){$t=64$ Myr}
    \end{picture}
    \caption{Logarithm of the projected density of cold ($T<2 \times 10^4$ K) gas at different times for simulations HD (top, no magnetic field), MHD-W (weak magnetic field) projected along $x$ (middle left) and $y$ (middle right), and MHD-S (strong magnetic field) projected along $x$ (bottom left) and $y$ (bottom right).}
    \label{fig:colddensproj}
\end{figure*}
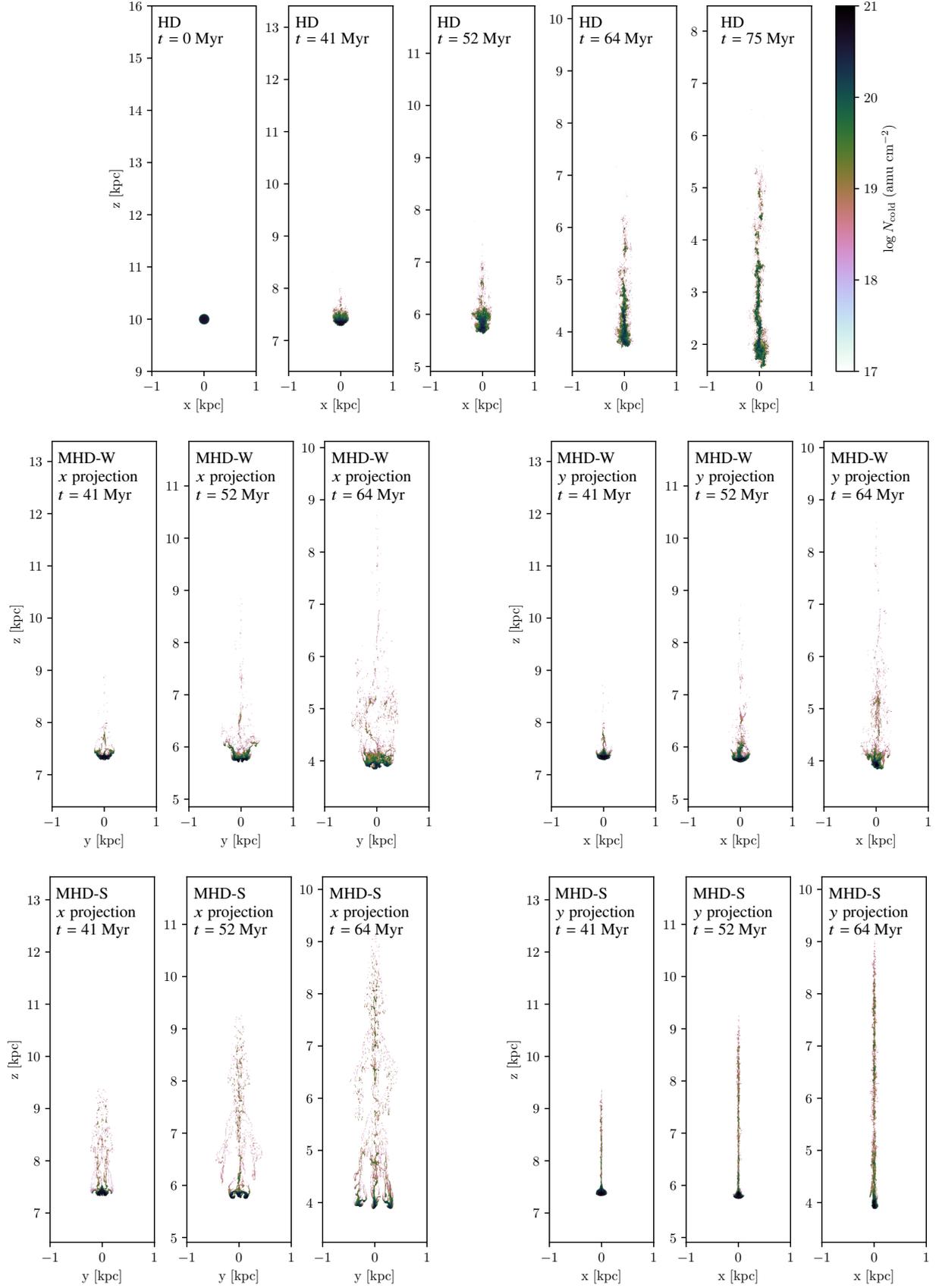

\subsection{Cloud evolution}
We show the projected density of cold gas at different times\footnote{Animated versions are available at \url{https://www.astro.rug.nl/~gronnow/animations/magnetichvcs.html}}
of the non-magnetic simulation, HD, weak field simulation MHD-W, and strong field simulation MHD-S, in Figure \ref{fig:colddensproj}.

In the non-magnetic case, the two transverse dimensions, $x$ and $y$, are similar and so we only show one projection. As can be seen, the evolution of the cloud's morphology in these simulations is clearly distinct. The evolution in the magnetic simulations is highly asymmetrical along the transverse dimensions, especially for the strong field case. The magnetic field stretches the cloud in the transverse direction perpendicular to it (i.e. along $y$) but for the strong field keeps the cloud very compact along it (i.e. along $x$). Both of the magnetic simulations show a large number of small, cold, and loosely connected cloudlets at late times while in simulation HD the head and tail remain mostly connected.

The strong field in simulation MHD-S exacerbates the RT-instability at the cloud's leading edge along $y$. There field lines become trapped in indentations and strongly amplified. Eventually, around $t\approx 45$ Myr this leads to the creation of finger-like structures which break apart and branch off forming a tree-like structure.

Later at $t\approx 55$ Myr another set of fingers are created, but these end up detaching from the main cloud completely, leading to the main cloud remnant splitting into three clumps as seen in the $x$-projection at $t=64$ Myr.

\begin{figure}
    \centering
    \includegraphics[height=0.4\textheight]{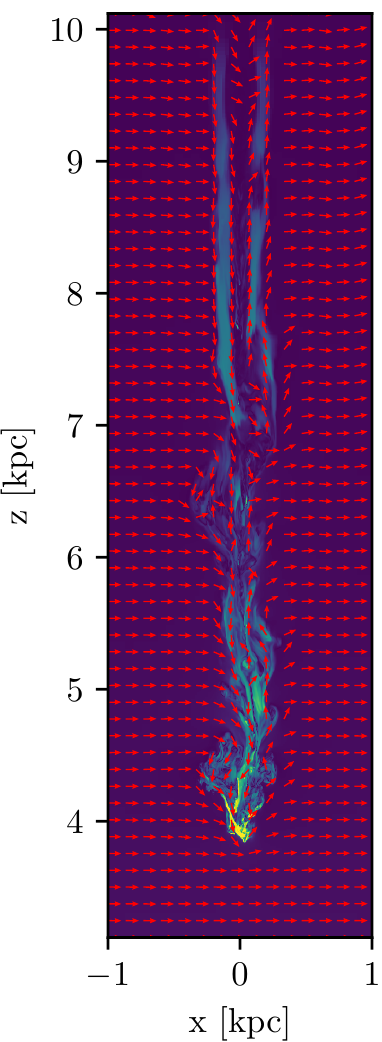}
    \includegraphics[height=0.4\textheight]{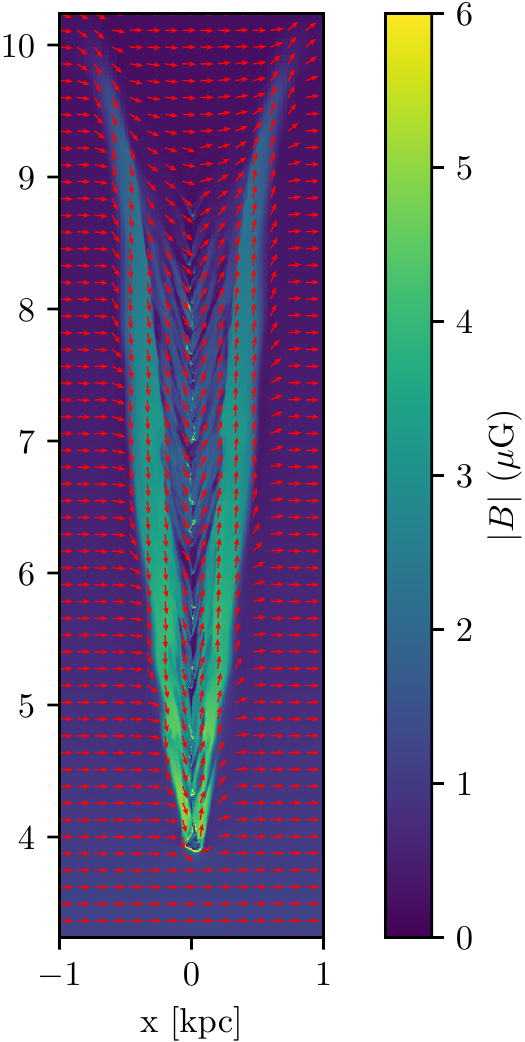}
    \caption{Magnetic field strength in a slice at $y=0$ at $t \approx 64$ Myr for simulation MHD-W (left) and MHD-S (right). Normalised arrows show the direction of the field in the $xz$-plane.}
    \label{fig:bmag}
\end{figure}

Figure \ref{fig:bmag} shows a slice of the magnetic field strength at $y=0$ for the two MHD simulations at $t=64$ Myr. As can be seen, the cloud strongly affects the magnetic field. More specifically, the field becomes locally strongly amplified and aligned with the cloud and wake. This is due to the magnetic `draping' effect as the cloud sweeps up the magnetic field as it moves through it \citep[see ][for a detailed description of this phenomenon]{dursi08,gronnow18}. In the wake of the cloud the draping causes oppositely directed magnetic field lines to come into close vicinity of each other. Hence, we expect that magnetic reconnection might occur in this region when resistivity is included.

\begin{figure}
    \centering
    \includegraphics[width=0.49\textwidth]{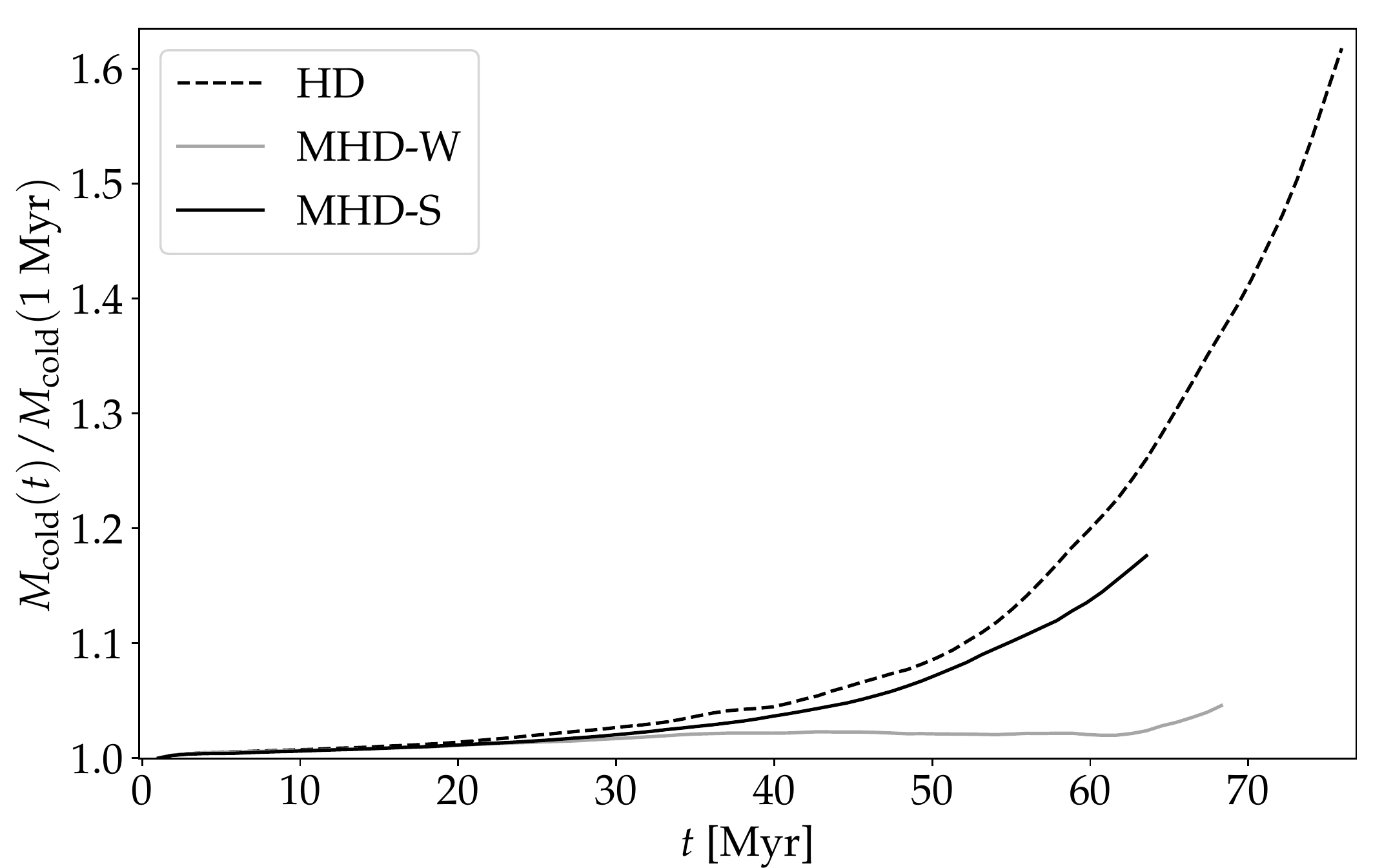}
    \caption{Cold gas mass evolution for the simulation without magnetic field (HD, dashed), with a weak field (MHD-W, solid grey), and a strong field (MHD-S, solid black). The vertical axis is the mass of gas at temperatures $T<2\times 10^4$ K divided by the mass of gas at these temperatures at $t=1$ Myr. This is to filter out the rapid early cooling which is not due to condensation as described in the text.}
    \label{fig:coldgas}
\end{figure}

\subsection{Condensation of cold gas}
\label{sec:condensation}
As mentioned previously, numerical studies in the literature have found that the mixing between cold clouds and the hot corona can cause cold gas to condense out of the hot gas. In this way, substantial amounts of cold gas might be accreted even if the original cloud has largely dispersed by the time it reaches the disc. This process is usually associated with relatively metal-rich intermediate-velocity clouds ejected in the galactic fountain \citep[e.g.][]{marinacci10,armillotta16,gronnow18,kooij21} but it has also been claimed to be efficient in metal-poor HVCs \citep{gritton17}.\\

We show the evolution of the total mass of cold ($T < 2\times 10^4$ K) gas in Figure \ref{fig:coldgas}. The cold gas mass is normalised by its value at $t\approx 1$ Myr. This is to filter out the rapid early cooling that occurs in the smooth transition region between the cloud and corona, i.e. at $r \approx r_c$. This cooling is not caused by mixing but rather by this gas having initial temperatures in between that of the cold gas and the hot corona near the peak of the cooling curve. As can be seen, in all cases the amount of cold gas is increasing with time, i.e. condensation is clearly occurring. The simulations that include the halo magnetic field both lead to less cold gas in agreement with the findings of \cite{gronnow18} for the case without any gravitational potential. However, the suppression in condensation is more effective for the \emph{weaker} magnetic field in simulation MHD-W where the mass of cold gas only increases appreciably after $t\approx 60$ Myr. A similar effect was noted in \cite{gronnow18} for the condensation of clouds in the galactic fountain. However, in that case this was caused by the stronger magnetic field not being effectively draped and amplified around the cloud. That cannot be the case here because significant draping is observed in both of the magnetic simulations. This is also expected because the cloud is moving faster than the Alfv\'{e}n speed at $t \gtrsim 25$ Myr in both cases and so is firmly in the regime where draping is effective. Instead, the cause for the greater mass of condensed gas in simulation MHD-S is the `fingers' created by RT-instability that are mostly absent in simulation MHD-W (see Figure \ref{fig:colddensproj}). These fingers split into cloudlets where mixing with hot gas is efficient leading to condensation. Of course, for magnetic fields progressively weaker than in MHD-W eventually the cold gas mass will tend towards the HD result \citep[see][]{gronnow18}.\\

\subsection{Cold gas velocity}
\label{sec:coldgasvel}
\begin{figure}
    \centering
    \includegraphics[width=0.49\textwidth]{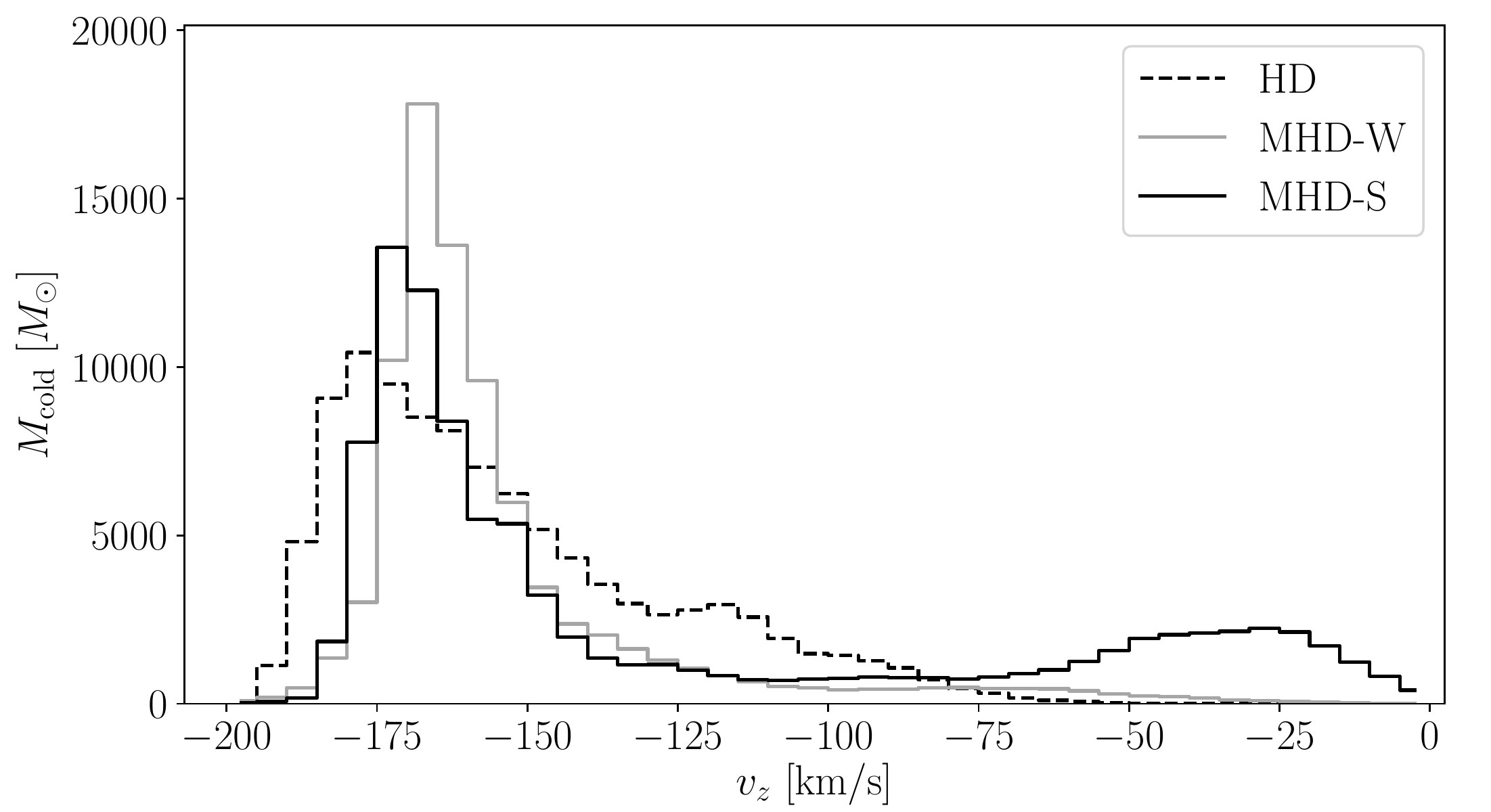}
    \caption{Velocity distribution (i.e. mass-weighted velocity histogram) of cold gas at $t=64$ Myr for the simulation without magnetic field (HD, dashed black), the weak magnetic field simulation (MHD-W, solid grey), and the strong magnetic field simulation (MHD-S, solid black).}
    \label{fig:coldgasvel}
\end{figure}

\begin{figure}
    \centering
    \includegraphics[width=0.49\textwidth]{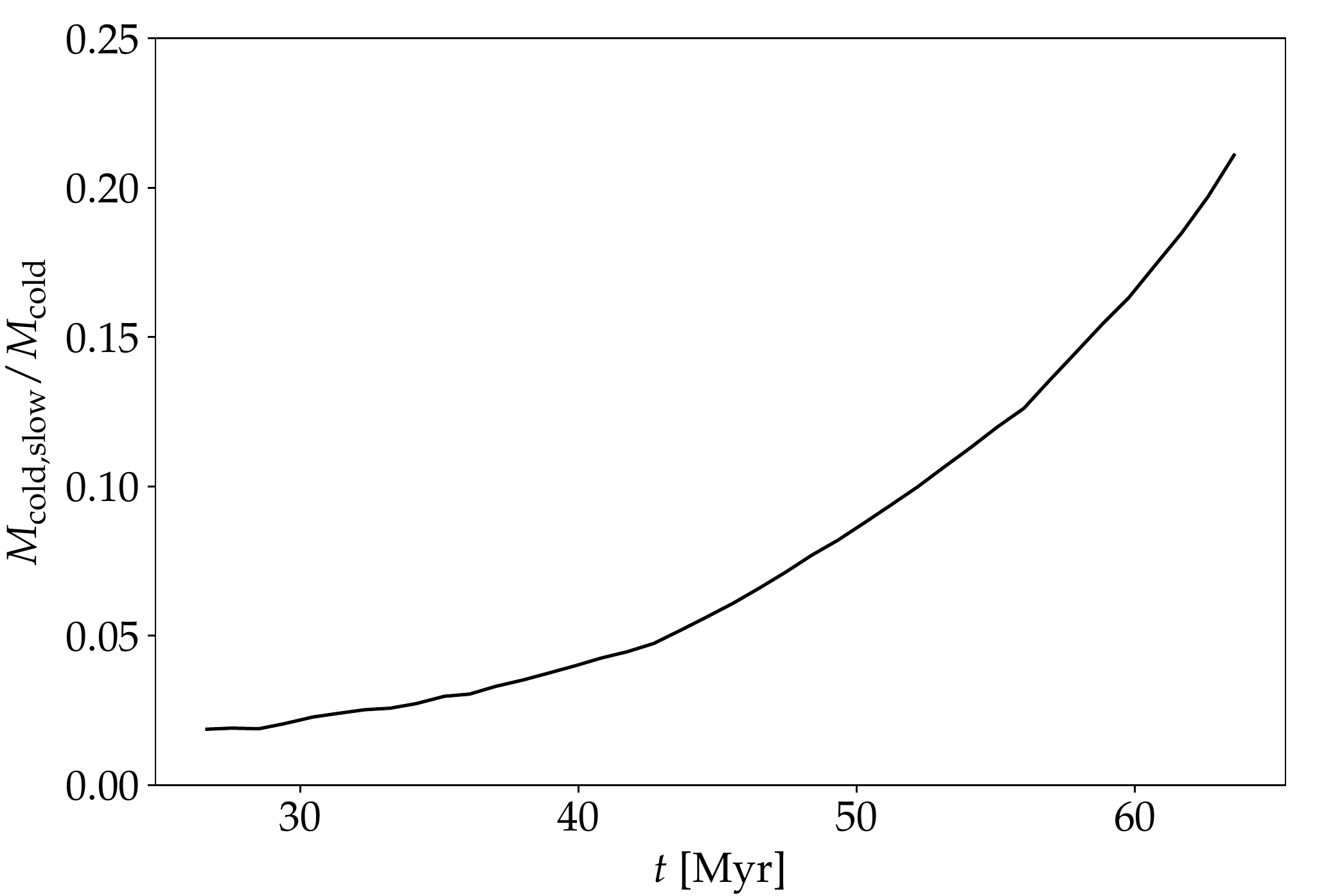}
    \caption{Fraction of cold gas mass at velocities $|v_z| < 60 \kms$ in simulation MHD-S after the main cloud remnant has accelerated beyond this velocity.}
    \label{fig:coldgasslowfrac}
\end{figure}

We show the velocity distribution of cold gas (calculated as the mass-weighted velocity histogram of cells containing cold gas) for our three standard simulations at $t=64$ Myr in Figure \ref{fig:coldgasvel}. As can be seen, in all cases, most of the cold gas reaches sufficient infall velocity to be classified as high-velocity (assuming that the velocity with respect to the corona is representative of the LSR velocity). However, the distributions at lower velocities\footnote{We use `lower' and `higher' for velocities based on the absolute velocity, i.e. $-200 \kms$ is a higher velocity than $-50 \kms$ despite being more negative, for example.} are clearly different. Simulation HD has a tail towards lower velocities due to interaction with the corona, as expected. However, for simulation MHD-S this tail is much less prominent and instead the velocity distribution is bimodal with a low-velocity bump peaking around $v_z = -30 \kms$. Overall, about 23 per cent of the total cold gas mass is in this low velocity bump at $|v_z| < 60 \kms$ (and about half of this below the peak at $|v_z| < 30 \kms$ as expected from its relative symmetry) in simulation MHD-S. We show the evolution of the fraction of cold gas that is at $|v_z| < 60 \kms$ in simulation MHD-S in Figure \ref{fig:coldgasslowfrac}. This fraction is rising superlinearly mirroring the evolution of the total cold gas mass in Figure \ref{fig:coldgas}. In contrast, without the magnetic field the velocity distribution essentially does not extend below $|v_z| \approx 60 \kms$ with less than 1 per cent of the cold gas mass being at those velocities. Simulation MHD-W has a less prominent tail towards lower velocities than simulation HD following the distribution of simulation MHD-S at velocities in between the two peaks. Unlike simulation MHD-S it has no clear second peak but it does have more gas at velocities $|v_z| < 75 \kms$ than simulation HD. At the end of simulation MHD-W at $t=68$ Myr the mass fraction of cold gas at $|v_z| < 60 \kms$ is about 4 per cent. Simulation HD continues to contain virtually no cold gas at $|v_z| < 60 \kms$ until it ends at $t=76$ Myr.

\begin{figure}
    \centering
    \includegraphics[width=0.49\textwidth]{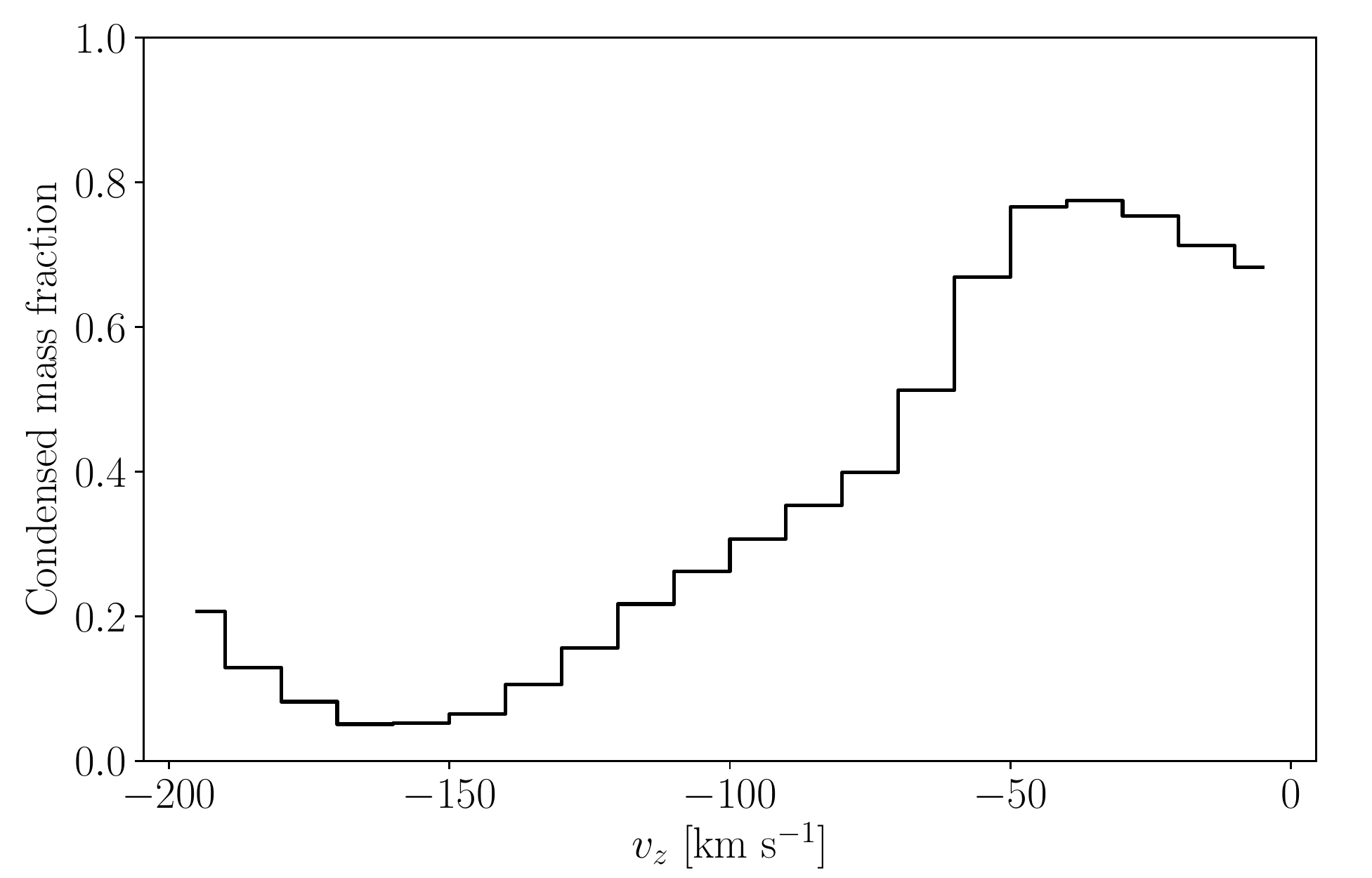}
    \caption{Fraction of condensed gas, i.e. mass fraction of cold gas that was not originally in the cloud according to the tracer, as function of velocity for simulation MHD-S at $t=64$ Myr.}
    \label{fig:coldgasfcond}
\end{figure}

In Figure \ref{fig:coldgasfcond} we show the velocity distribution of the condensed mass fraction of cold gas in simulation MHD-S at $t=64$ Myr. This is the fraction of cold gas in each velocity bin that did not originally belong to the cold cloud, as tracked by the passive scalar $C$ (see Section \ref{sec:nummethod}). Hence, it represents gas that was initially in the hot corona and has cooled during the simulation due to mixing with cloud material (i.e. condensed). As expected, this reveals that the cold gas in the high-velocity peak largely represents the main remnant(s) of the original cloud which have become HVCs. Conversely, the low velocity bump is largely comprised of gas that has condensed during the simulation. This is consistent with the increase over time of slow gas seen in Figure \ref{fig:coldgasslowfrac}. The fact that the lowest velocity gas is largely comprised of the initially static coronal gas is of course expected in any case. However, when the magnetic field is absent this condensed gas is able to accelerate to, or at least maintain, intermediate and high velocities and form a tail rather than a second peak. We interpret this as being due to the effect of magnetic tension from strongly draped fields as described in Section \ref{sec:slowclumps}. For the weak field, there is a low fraction of gas at velocities below the main peak due to the low amount of condensation.

\subsection{Clump finding analysis}
\label{sec:clumps}

\begin{figure*}
    \centering
    \includegraphics[width=\textwidth]{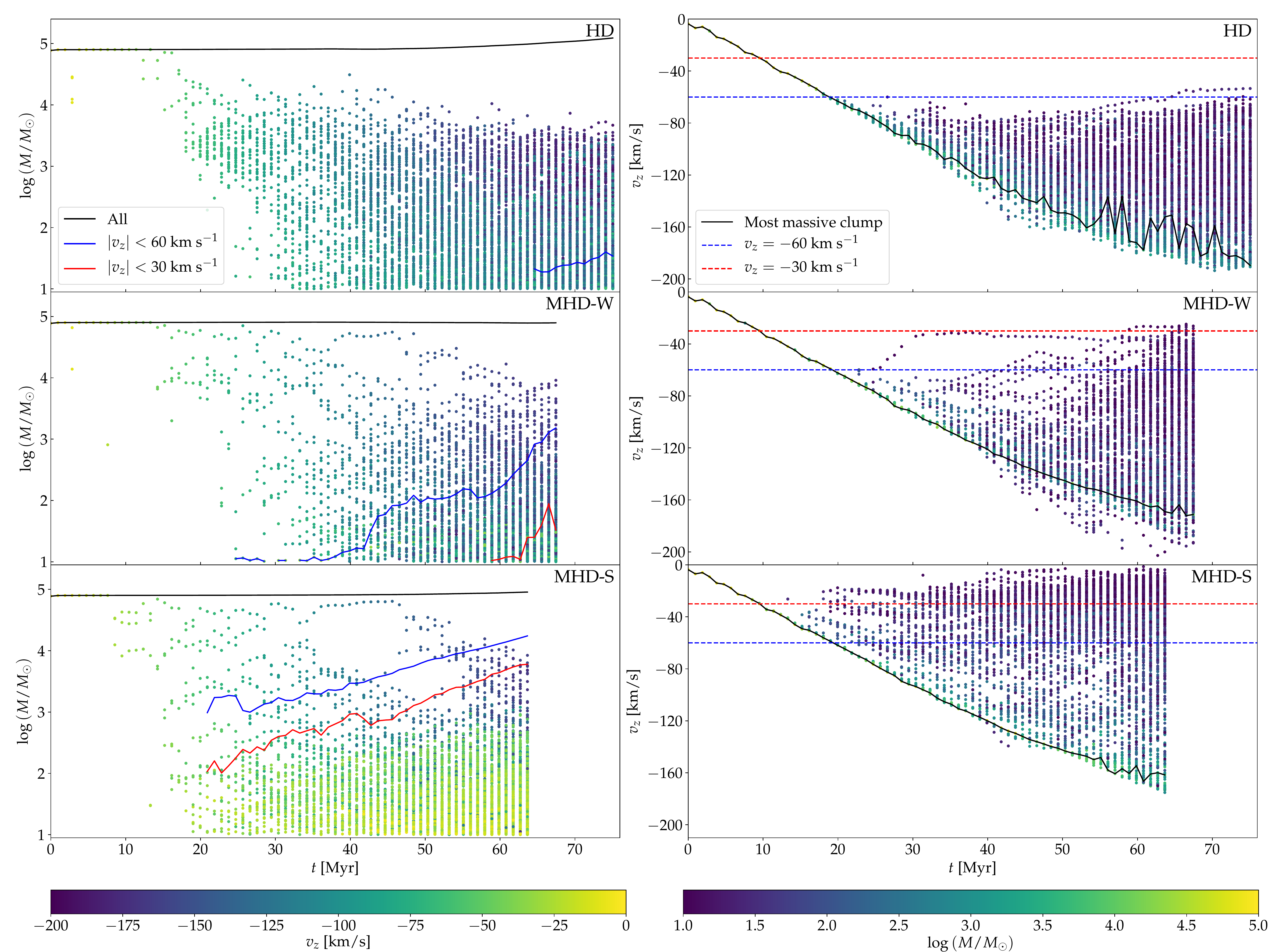}
    \caption{The mass and velocity of clumps as a function of time, for simulations HD (top), MHD-W (middle), and MHD-S (bottom). On the left, mass is on the $y$-axis while the velocity is indicated by the colour, and on the right this is swapped. On the left, the black curve indicates the total mass of clumps, the blue curve indicates the mass of low-intermediate velocity clumps ($|v_z| < 60 \kms$), and the red curve indicates the mass of low velocity clumps ($|v_z| < 30 \kms$). On the right, the black curve indicates the velocity of the most massive clump, i.e. the main remnant of the HVC, while the dashed blue and red lines mark the $60 \kms$ and $30 \kms$ thresholds, respectively.}
    \label{fig:clumps}
\end{figure*}
\begin{figure}
    \centering
    \includegraphics[width=0.41\textwidth]{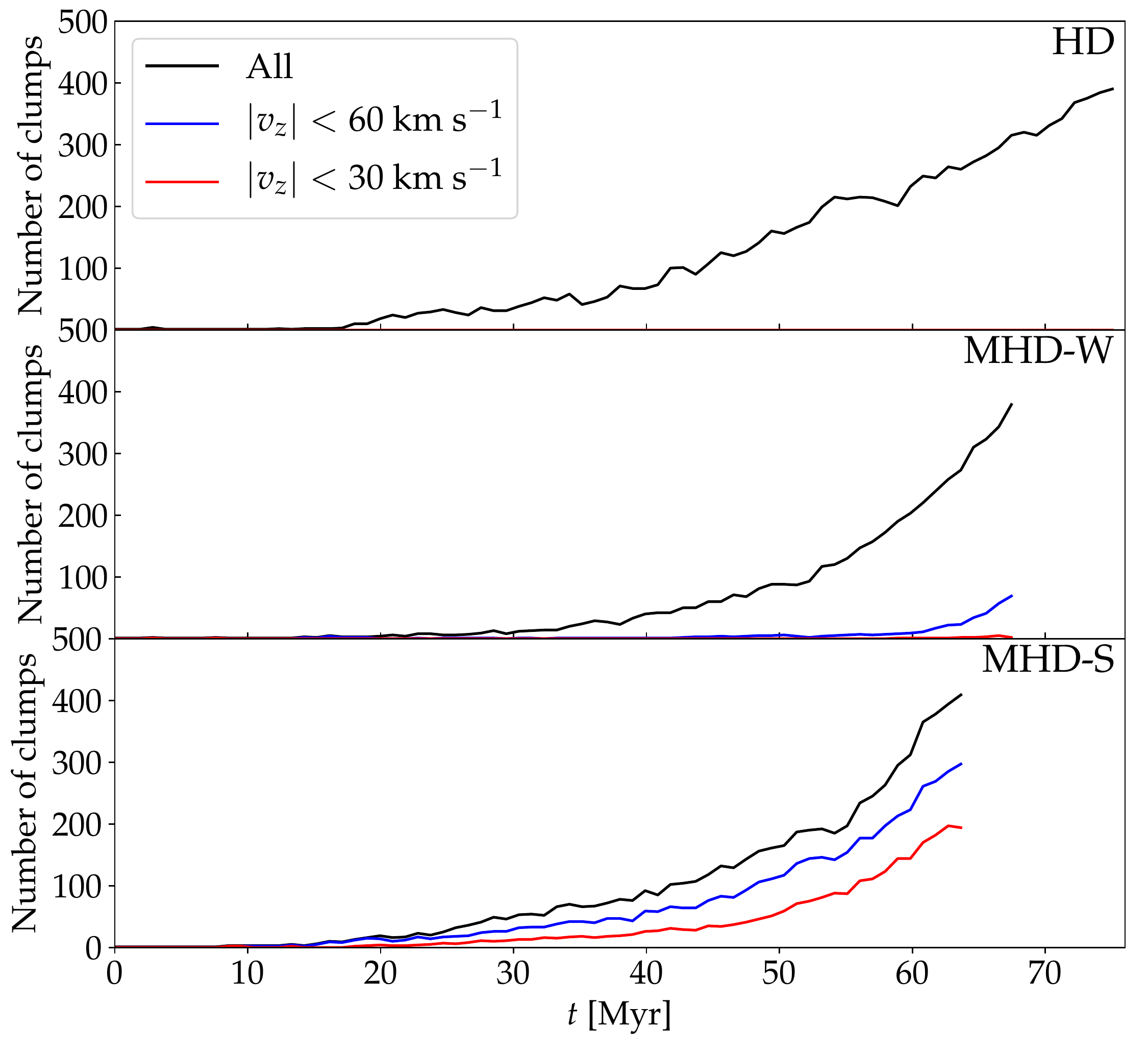}
    \caption{The total number of detected clumps (black), low-intermediate velocity clumps ($|v_z| < 60 \kms$, blue) and low velocity clumps ($|v_z| < 30 \kms$, red) in simulations HD (top), MHD-W (middle), and MHD-S (bottom).}
    \label{fig:nclumps}
\end{figure}

From Figure \ref{fig:colddensproj} it is clear that in all cases the cloud breaks into clumps and filaments. As discussed in \ref{sec:condensation}, the stripped gas is also not smoothly distributed because it triggers condensation of coronal gas creating cloudlets that grow through cooling. Therefore, in addition to examining the integrated mass, we ran a clump finding algorithm to study the mass and velocity spectrum of these clumps. We used the FellWalker \citep{berry15} algorithm as implemented in CUPID \citep{berry07} on the density outputs from the simulations. To avoid spurious density peaks being interpreted as clumps we required a 48 cell minimum for each clump as well as a 10 $M_\odot$ mass minimum. We did not put any constraint on the temperature of the gas as any significant overdensities in our simulations are cold anyway. We have verified that we find essentially the same spectrum of clumps when running the clump finder on the density of only the cold ($T < 2 \times 10^4$ K) gas instead. Hence, these clumps are tracking the cold gas distribution.

We show the mass and centre of mass z-velocity distribution of these clumps for simulations HD, MHD-W, and MHD-S, as a function of time in Figure \ref{fig:clumps}. In all simulations the most massive clump (i.e. the main HVC remnant) moves with constant acceleration towards the disc until roughly $t\approx 50$ Myr. After this time, drag becomes significant and in all simulations the velocity of the main remnant appears to have flattened out by the end of the simulation. The increase in the total mass in clumps is due to more clumps forming through condensation rather than individual clumps gaining mass. This is clear from Figures \ref{fig:clumps} and \ref{fig:nclumps}. Figure \ref{fig:clumps} shows that the clumps, including the main remnant, are generally losing mass. Figure \ref{fig:nclumps} shows that the number of clumps increases substantially with time.

Examining the rest of the distribution, it is clear that a large population of clumps with low masses and low velocities exist for MHD-S which is absent in the HD case. These clumps make up the low velocity bump seen in the overall velocity distribution of the cold gas as previously described in Section \ref{sec:coldgasvel}. In simulation HD, at all times all clumps, including ones with low mass, have velocities of $|v_z| > 50 \kms$ as expected from the cold gas velocity distribution. In contrast, most of the low mass ($M < 100 M_\odot$) clumps have velocities below this value throughout most of the strong field simulation. The weak field case is intermediate between the two with some low velocity clumps. The velocity of the main cloud remnant, i.e. the most massive clump, as shown by the solid black line in the velocity plot in Figure \ref{fig:clumps} (right panel), is slightly lower at late times when the magnetic field is included. However, this difference is much smaller than the general difference in the velocities of the lower mass clumps.

We classify clumps with $|v_z| < 60 \kms$ as `low-intermediate velocity' clumps (LIVC) and clumps with $|v_z| < 30 \kms$ as `low velocity' clumps (LVC) indicated by the dashed blue and red lines in the velocity plot, respectively. These velocities correspond to the upper bound of the low velocity bump in simulation MHD-S at $t=64$ Myr (see Figure \ref{fig:coldgasvel}) and to approximately the regime for LSR velocity measurements usually considered to be low-velocity gas, respectively. We show the integrated mass of these clumps with the solid blue and red lines in the mass plot in Figure \ref{fig:clumps}. As can be seen, this mass is low compared to the total mass of all clumps, which is dominated by the main cloud remnant, but it is growing much faster than the overall mass in clumps. The fraction of mass in LIVCs and LVCs is about 20 and 7 per cent, respectively, at the end of simulation MHD-S at $t\approx 64$ Myr. This is a bit less than the overall fractions of the total cold gas at those velocities (see Section \ref{sec:coldgasvel}) so some of this mass is presumably somewhat diffuse in the tails of clumps and filaments between them at densities too low to be recognised as belonging to a clump by the clump finding algorithm. The mass fraction of LIVCs for simulation MHD-W and also of LVCs for simulation MHD-S is increasing fast by the end of the simulations so we expect both to become a significant contribution by the time the main remnants would reach the disc-corona interface at $t\approx 75$ Myr. In any case, due to their growing importance with time as well as resolution (see Section \ref{sec:resolution}) the mass fraction in LIVCs and LVCs in our simulations is a lower limit. As expected from our results for the overall cold gas in Section \ref{sec:coldgasvel}, LIVCs and LVCs are mainly comprised of cold gas that has condensed out of the corona.

\begin{figure}
    \centering
    \includegraphics[width=0.36\textwidth]{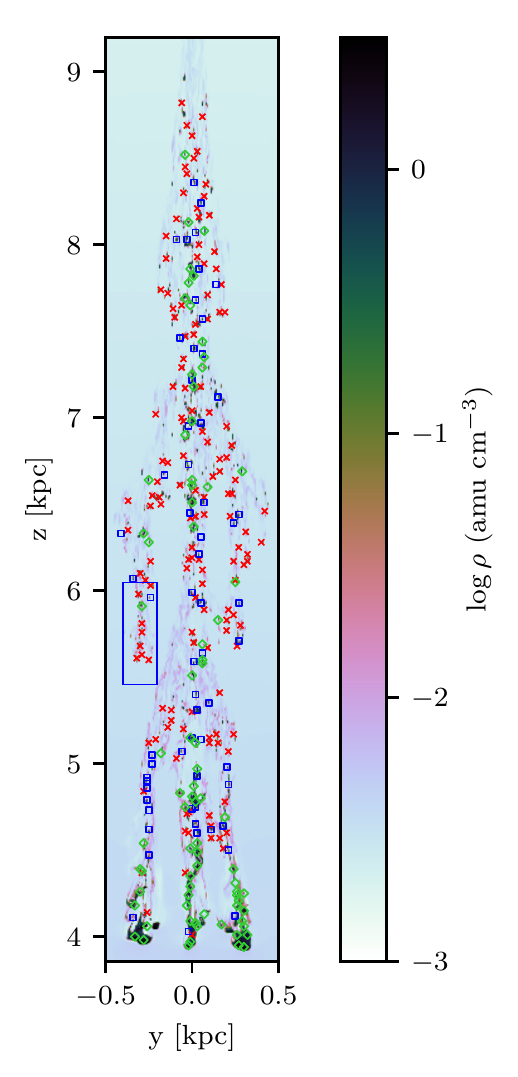}
    \caption{Logarithm of the density in a slice at $x=0$ at $t = 64$ Myr for simulation MHD-S. The centre of mass positions of clumps in this slice are indicated with different symbols according to their velocities. These are red crosses for $|v_z| < 30 \kms$, blue squares for $30 \kms < |v_z| < 60 \kms$, and green diamonds for $|v_z| > 60 \kms$. The blue rectangle indicates the region plotted in Figure \ref{fig:clumpvel}. The slice in the $xz$ plane in Figure \ref{fig:clumpfield} is centered on the middle of this rectangle at $y=-0.3$.}
    \label{fig:clumplocs}
\end{figure}

We show the position of the clumps distinguished by their velocities on top of the density at $x=0$ at the end of simulation MHD-S at $t=64$ Myr in Figure \ref{fig:clumplocs}. As can be seen, LVCs and LIVCs generally reside in the wake while the faster clumps are mostly near the front, close to the main cloud remnants.

\section{Discussion}
\label{sec:discussion}
\subsection{Magnetic deceleration of condensed gas}
\label{sec:slowclumps}
\begin{figure}
    \centering
    \includegraphics[width=0.49\textwidth]{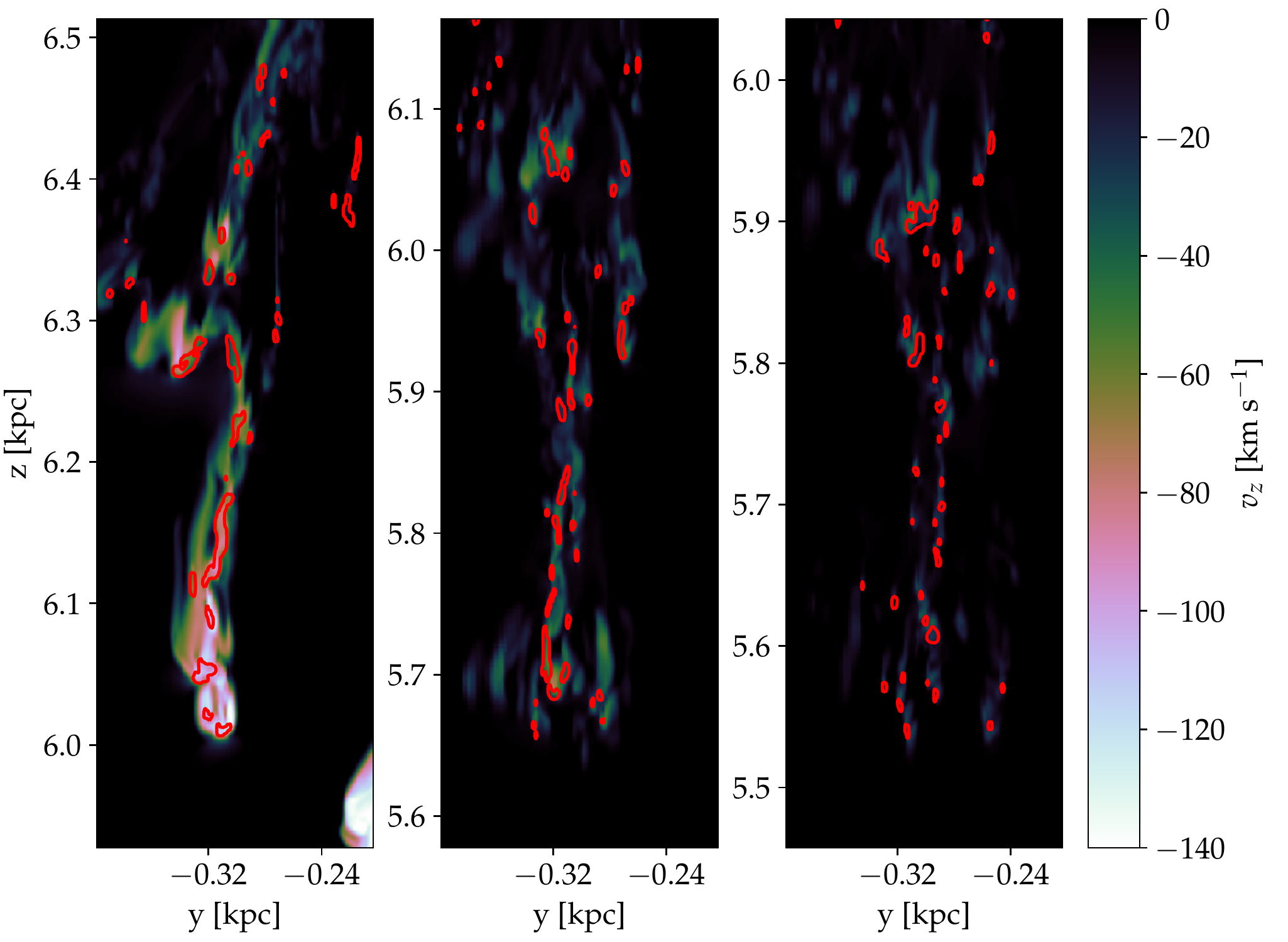}
    \caption{Velocity along $z$ in a slice at $x=0$ of the small region with condensed cloudlets marked in Figure \ref{fig:clumplocs} in simulation MHD-S at $t=52$ Myr, $t=60$ Myr, and $t=68$ Myr from left to right, respectively. Overlaid in red are density contours at $\rho=0.2 \pcc$. The filament breaks apart and decelerates.}
    \label{fig:clumpvel}
\end{figure}

\begin{figure}
    \centering
    \includegraphics[width=0.32\textwidth]{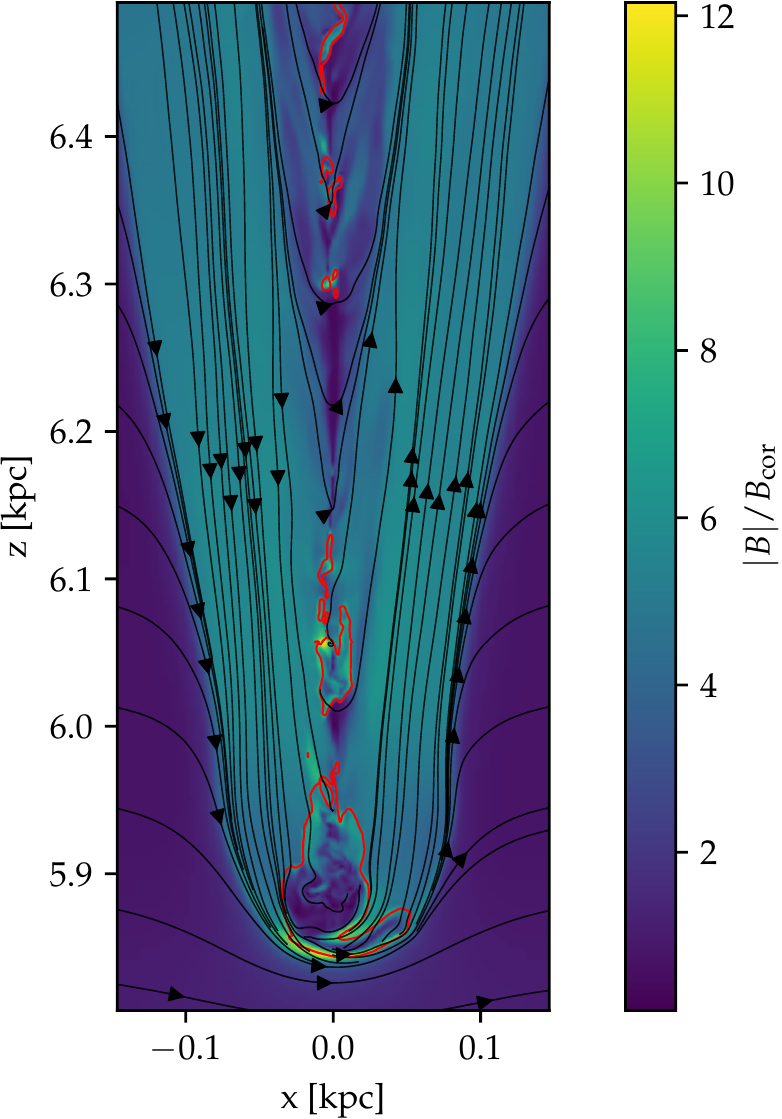}
    \caption{Magnetic field strength divided by the coronal field strength in a slice at $y=-0.3$ of the small region with condensed cloudlets marked in Figure \ref{fig:clumplocs} in simulation MHD-S at $t=52$ Myr. Overlaid in red are density contours at $\rho=0.2 \pcc$ and magnetic field lines in this plane.}
    \label{fig:clumpfield}
\end{figure}

As shown in Sections \ref{sec:condensation} and \ref{sec:clumps}, much of the condensed cold gas is severely decelerated in simulation MHD-S. We show the velocity evolution of some of these clumps in one of the finger-like substructures in Figure \ref{fig:clumpvel}. The region shown in this figure is marked by the blue rectangle in Figure \ref{fig:clumplocs}. Clearly, the cold gas in this structure is strongly decelerating. The cloudlets there might be more affected by hydrodynamic drag due to not being directly upwind of the main cloud remnants. However, more importantly is the deceleration caused by the magnetic field. The Lorentz force of the magnetic field is given by
\begin{equation}
\mathbf{f}_L=\frac{1}{4\pi}(\nabla\times\mathbf{B})\times\mathbf{B}=-\frac{1}{8\pi}\nabla B^2+\frac{1}{4\pi}(\mathbf{B}\cdot\nabla)\mathbf{B},
\end{equation}
where the first term is the force due to magnetic pressure and the second term is the force due to magnetic tension. The magnetic tension force resists curving of field lines and can be written as
\begin{equation}
\mathbf{f}_t = \frac{1}{8\pi}\mathbf{\hat{b}}\mathbf{\hat{b}}\cdot\nabla B^2 + \frac{B^2}{4\pi R_c}\mathbf{\hat{n}},
\end{equation}
where $\mathbf{\hat{b}}$ is the unit vector in the direction of $\mathbf{B}$, $\mathbf{\hat{n}}$ is the unit vector perpendicular to the field lines, and $R_c$ is the radius of curvature of the field lines. Figure \ref{fig:clumpfield} shows that the field is strongly draped in the $xz$-plane around the structure shown in Figure \ref{fig:clumpvel}. The draping causes amplification of the field in a layer bending around the front of the clumps. This leads to a force from the magnetic pressure in the $-z$ direction at the front. Additionally, the strong bending of the field lines in this layer leads to strong magnetic tension. At the front of draped clumps this force likewise points in the $-z$ direction. Hence, magnetic pressure and tension forces effectively act as additional drag. We find that the magnetic tension force generally dominates over the force due to magnetic pressure in draped layers by typically roughly an order of magnitude. Hence, magnetic tension is the main cause of the deceleration.

The overall effect of magnetic fields decelerating gas clouds is known from previous wind tunnel simulations \citep[e.g.][]{gronnow18,gronke20a,sparre20}, as well as cosmological zoom-in simulations \citep{vandevoort21} and the largely similar MHD hydrostatic corona setup of \cite{kwak09}. However, {\em we uncover an important aspect of this effect in our falling cloud simulations due to the interplay of magnetic fields, the gravitational potential, and radiative cooling not seen in previous studies} which did not include all these effects: The magnetic drag does not affect all the gas in our simulations equally. The clumps that are either a remnant of the original HVC or were stripped at early times are able to accelerate largely unimpeded by the magnetic field. Eventually they build up a considerably amplified draped field at their leading edge but have sufficient momentum to not be significantly decelerated by this. In contrast, the stripped gas, where the mixing and condensation occurs, is decelerating in some cases to the point of being nearly suspended in the corona. These cloudlets have lower momentum and additionally, due to their small sizes, the radius of curvature of the draped field around them is small compared to around the main cloud remnants leading to much stronger tension force. This leads to the overall bimodal velocity distribution with a low-velocity bump, rather than a tail, as shown in Section \ref{sec:condensation}.

Recently, \cite{heitsch21} performed simulations with a setup reminiscent of ours but without a magnetic field, and found that condensed gas was able to catch up to, and in some cases even overtake, the original cloud material in a `peloton effect'. In Figure \ref{fig:clumps} it can be seen that there are some clumps at late times that are travelling slightly faster than the most massive cloud remnant in simulations HD and MHD-W. This indicates that the peloton effect might be occurring to some degree in these cases. However, the magnetic deceleration of condensed gas in simulation MHD-S appears to suppress this effect almost completely. Hence, the peloton effect might not be important in the inner part of the corona but could still be effective at larger distances where the halo magnetic field is weaker.

\subsection{Efficiency of accretion through condensation from HVCs}
We have shown that the halo magnetic field substantially decreases the efficiency of condensation of cold gas around an HVC. As discussed in \cite{gronnow18} the magnetic field will be amplified and bent around the clouds. If the cloud is traveling faster than the Alfv\'{e}n speed in the hot gas the field becomes completely draped around the cloud and its wake. This is obviously not the case at early times as the cloud is initially comoving with the corona, but it eventually happens in all cases as the cloud accelerates. The amplification and the alignment of the field with the interface caused by the draping suppresses the Kelvin-Helmholtz instability. This inhibits the mixing of cold cloud and hot coronal gas which drives the condensation.

In addition, we also expect the significant deceleration of some of the condensed cloudlets in a moderately strong field described previously to further slightly restrict the accretion of condensed gas. Due to their slow velocities, the time it takes them to reach the disc becomes hundreds of Myr. Rather than delaying the accretion of these cloudlets, they might not be accreted at all. Because of their small sizes and long infall times they may be dispersed by thermal conduction before they can reach the disc. However, once the clumps have decelerated to very low velocities they no longer effectively sweep up surrounding field lines. The ambient magnetic field by itself might not provide sufficient tension, in which case the time needed to reach the disc for these clumps will depend on how long the magnetic field can remain locally draped and amplified before decaying from magnetic diffusion processes. The strong local magnetic field will also strongly suppress thermal conduction. However, the draped field will be locally ordered, even if the surrounding field has a significant random component, and thermal conduction can still be non-negligible in that case \citep{kooij21}. Future simulations that include anisotropic thermal conduction as well as non-ideal MHD effects are needed to further examine this.

In spite of these effects, the amount of gas that can be accreted through the condensation triggered by HVCs remains substantial. As can be seen from Figure \ref{fig:coldgas}, the condensation increases the mass of cold gas by almost 20 percent by 64 Myr for simulation MHD-S and at least half of this is intermediate velocity gas. While the weaker magnetic field of simulation MHD-W is much more effective at suppressing condensation by not creating the RT fingers seen in MHD-S, cold gas mass is increasing at late times and the condensed cloudlets are being decelerated much less and can hence more easily reach the disc. This agrees with the simpler HVC simulations without magnetic fields or a gravitational potential of \cite{gritton17}. However, we expect that we are overestimating the amount of condensed gas during the simulations because we do not include heating from ultraviolet (UV) radiation and cosmic rays and ignore thermal conduction which will reduce the efficiency of condensation \citep[][see Section \ref{sec:limitations}]{kooij21}. Nonetheless, this suggests that the gas accretion rate from HVCs might be somewhat underestimated in the literature \citep[e.g.][]{putman12} although by an amount still within the substantial uncertainties of such studies. A larger suite of simulations that explore the parameter space of initial conditions (e.g. clouds with different initial sizes and $z$ coordinates or different trajectories) would be needed to quantify the overall importance of additional accretion through condensation.

\subsection{Observational detection of magnetically decelerated gas}
Clearly, the low velocity clumps are too small to be detected individually. In theory, the overall bimodal velocity distribution with the low velocity bump of largely condensed gas seen in simulation MHD-S (see Figure \ref{fig:coldgasvel}) should be discernible for HVCs traveling through parts of the corona where the halo magnetic field is moderately strong. For this, the observed infalling clouds would have to be at high galactic latitude where the line of sight is largely aligned with the cloud's velocity \citep[e.g.][]{bish19}. However, in practice gas at these low velocities would be largely indistinguishable from the foreground gas in the interstellar medium (ISM) in the disc. In some cases, the velocity distribution of the low velocity bump might be partially separated from that of the ISM because it is centered at a slightly negative velocity rather than at zero. However, it would generally not be possible to establish if such a peak represents a tail of trailing condensed material or an unrelated component somewhere along the line of sight. A precise measurement of the chemical composition could assist in this although, being mostly condensed gas, we would expect the composition to be close to that of the corona.

\subsection{Resolution}
\label{sec:resolution}
From previous studies of both wind tunnel and hydrostatic simulations of clouds we expect our standard resolution of 50 cells per cloud radius (which we denote as $\mathcal{R}=50$) to be sufficient to capture the overall qualitative evolution. However, we also expect it to be insufficient for convergence of some quantities, especially those that depend on the cooling such as total cold gas mass. For adiabatic MHD simulations resolutions in the range 32-64 cells per cloud radius have been found to be needed \citep{dursi08,banda-barragan16}. For radiative cooling the physical cell size matters. Without thermal conduction, the cooling will tend to break up cloudlets all the way down to the size where the sound crossing time becomes comparable to the cooling time scale \citep[refered to as `shattering', see e.g.][]{mccourt18,gronke20b,sparre20}. This scale is typically unfeasibly small to resolve for simulations of Galactic gas clouds and is far from resolved at the $\Delta x=2$ pc standard resolution (i.e. $\mathcal{R}=50$) in our simulations. Hence while the main cloud remnant should be relatively well resolved, the broad spectrum of stripped and condensed clumps will not be.

\begin{figure}
    \centering
    \includegraphics[width=0.49\textwidth]{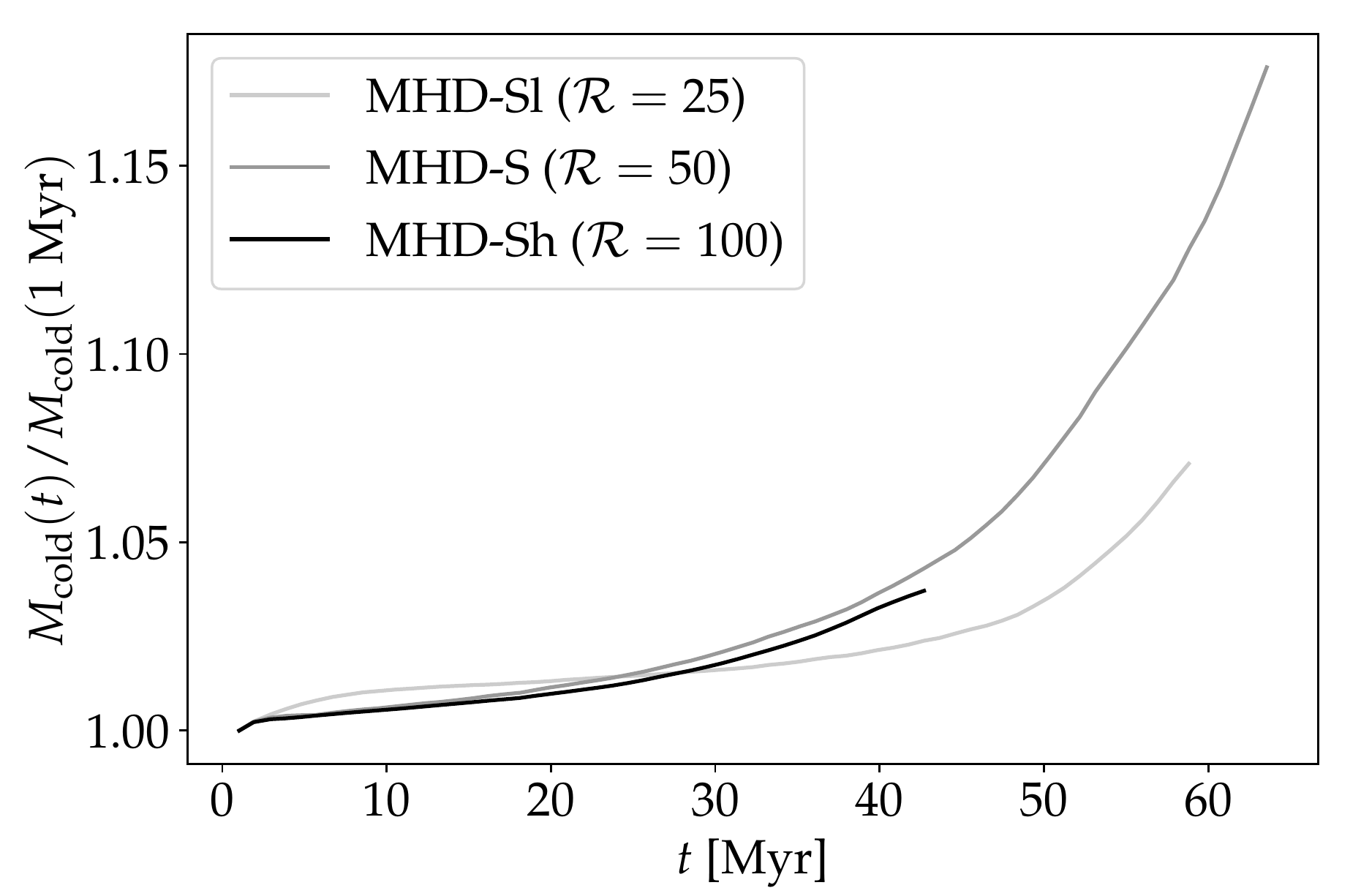}
    \caption{Cold gas mass evolution for the simulation with a strong magnetic field at low resolution (light grey), standard resolution (dark grey), and high resolution (black).}
    \label{fig:convergence}
\end{figure}

In Figure \ref{fig:convergence} we show the cold gas mass evolution for simulation MHD-S at three different resolutions, the standard resolution ($\mathcal{R}=50$) and half and twice the resolution ($\mathcal{R}=25$ (MHD-Sl) and $\mathcal{R}=100$ (MHD-Sh), respectively). The lowest resolution run clearly underestimates the mass of cold gas, except at early times where the higher numerical diffusion at the cloud-corona interface in this simulation leads to too much cooling. Due to its high computational cost, we could only run the highest resolution simulation to $t=42$ Myr. It closely follows the standard resolution case deviating only slightly towards lower masses at $t\gtrsim 25$ Myr. While MHD-Sh ends too early for much of the `tree-like' structure to have formed, clumps have started condensing in the wake and MHD-Sl is already significantly diverging from MHD-S and MHD-Sh by this time. Running the clump finding algorithm on the outputs of simulation MHD-Sh at full resolution is not feasible. However, downsampling it by one AMR level, bringing it to the same resolution as MHD-S, we find that about a factor of two more clumps with generally lower masses are detected. If we lower the $10 M_\odot$ minimum mass to $1 M_\odot$ this difference disappears. Crucially, the mass fraction of cold gas that is in clumps with $|v_z| < 60 \kms$ (i.e. LVCs and LIVCs) is quite close at all times, being slightly higher for MHD-Sh. On the other hand, as expected, simulation MHD-Sl, which underestimates the overall mass of cold gas, correspondingly underestimates the number of clumps and greatly underestimates the mass fraction of LVCs and LIVCs.

In summary, the mass of cold gas as well as the mass fraction in LVCs and LIVCs appear to be quite well converged at our standard resolution of $\mathcal{R}=50$, although we cannot assess this at late times. In any case, a resolution not much lower than $\mathcal{R}=50$ is needed as our $\mathcal{R}=25$ simulation clearly disagrees with the results at higher resolutions. Based on this, we conclude that the condensation, and in particular the population of low-velocity condensed clumps, appears to be real rather than an artefact caused by insufficient resolution.

\subsection{Limitations and missing physics}
\label{sec:limitations}

The properties of individual clumps and the total number of clumps are sensitive to the choice of the clump finding algorithm and the parameters passed to it. Based on Figure \ref{fig:clumplocs}, Fellwalker \citep{berry15} appears to have identified clumps to good accuracy without counting single structures as multiple clumps or missing significant structures. To assess the robustness of the results of our clump based analysis in more detail, we have also run the CLUMPFIND algorithm \citep{williams94} as implemented in CUPID. In this case we likewise find a comparable population of slow, low mass clumps in simulation MHD-S that is absent in simulation HD.

We neglect heating and ionisation from UV photons in our simulations. This effect is highly uncertain in the part of the corona that we simulate. Both the metagalactic UV background and stellar UV radiation are likely significant and their relative importance changes with the height above the disc. We are, however, effectively assuming that heating dominates at low temperatures due to our $10^4$ K cooling floor. In general, UV heating will reduce the peak of the cooling curve around $T\sim 10^5$ K and lead to less condensation. On the other hand, if heating is included instead of a cooling floor, dense clumps can cool to temperatures below $10^4$ K and be less easily dispersed.

In our simulations the HVC travels perpendicular to the magnetic field lines. In reality the HVC would also experience a gravitational pull along $R$ and the ordered component of the field would not be purely perpendicular to $z$ \cite[e.g. there are indications of an `X-shaped' component of the halo magnetic field][]{jansson12a}. The interaction of clouds with uniform magnetic fields with different orientations has been studied in wind tunnel simulations. These find that while the effect of a field parallel to the wind is qualitatively different from a transverse field, in the general case of a field at some intermediate angle the evolution qualitatively follows the transverse case even for quite shallow angles \citep[see e.g. ][]{banda-barragan16,gronnow18,kooij21}. An additional complication in the Galactic magnetic field structure that we ignore is its random components \citep{jansson12b,beck16}. Turbulence adds isotropic and anisotropic random components to the halo magnetic field but this is very poorly constrained. In general, we expect any random field to become locally ordered around the cloud as the transverse components are draped as seen in simulations that include such fields \citep{asai07,sparre20}. Hence, including the random halo field should not fundamentally change our results.

We have ignored thermal conduction in our simulations, which may be important for the evolution of the cold gas. Magnetic fields strongly suppress this effect perpendicular to the field lines, however due to the field becoming locally ordered thermal conduction will generally not be completely suppressed along all directions. This anisotropic suppression has recently been found to be roughly equivalent to an overall isotropic decrease in thermal conduction efficiency by a factor of order 10 for the cold gas evolution in galactic fountain clouds \citep{kooij21}. However, this factor might be different for the environment of our simulations. Thermal conduction tends to lower stripping and disperse smaller cloudlets \citep{armillotta16,armillotta17}. Thus, as previously mentioned, the slow low mass clumps in the magnetic simulations would probably not survive to be accreted.

Kinematic models constrained by O\textsc{vii} absorption in the Milky Way's corona suggest that it is rotating at a significant velocity of $v_{\phi}=183\pm 41 \kms$ \citep{hodges-kluck16}, similar to theoretical expectations \citep{pezzulli16,pezzulli17}. The inclusion of rotation would change the density of the hydrostatic corona in our simulations. For the non-magnetised case the hydrostatic equilibrium solutions for such rotating coronae can be described with a relatively simple model \citep{sormani18}. However, this is not possible in general when a non-uniform magnetic field is included. In any case, the density difference between models with and without rotation is generally within the significant uncertainties in the density normalisation $n_{h,0}$ of non-rotating coronae. Additionally, the rotation velocity is expected to change with $z$ \citep{sormani18} which would lead to additional shear between the corona and the cloud.

We neglect the self-gravity of the gas. \cite{li20} explored the effect of self-gravity in clouds with similar flat density profiles as ours and found that for clouds initially below the Jeans mass self-gravity continues to be unimportant throughout the cloud's entire evolution. The mass of our cloud is an order of magnitude below its Jeans mass. Recently, however, \cite{sander19,sander21} showed that self-gravity can have a significant effect on clouds below the Jeans mass if they have a cuspy density profile. In that case, the self-gravity inhibits stripping such that the cloud loses gas more slowly but also decreases the efficiency of condensation lowering the overall amount of cold gas.

In reality observed HVCs are not described by a single flat or cuspy density profile but have structure on a wide array of scales. Cold gas with an initially fractal structure has been simulated in the context of the Magellanic Stream by \cite{bland-hawthorn07,tepper-garcia15}. In general, internal velocities and magnetic field configurations in the cloud will also be non-uniform. For such non-uniform clouds the magnetic field can be more effectively folded around the many substructures and have a stronger effect on the evolution \citep{banda-barragan18}. The evolution of initially non-uniform HVCs is outside the scope of this paper and we defer this work to a forthcoming paper (Jung et al., in prep).

\section{Summary}
\label{sec:summary}
We follow the evolution of a cloud formed out of the Galactic corona as it falls towards the disc due to its gravitational potential eventually becoming an HVC, with and without including the Galactic halo magnetic field. Summarising our main results, we find that:

\begin{enumerate}
    \item Although we are unable to follow the cloud all the way to the disc-corona interface when we include the magnetic field, the original cloud appears to survive. However, when a strong magnetic field is included the cloud breaks into multiple fragments due to enhanced Rayleigh-Taylor instability in the direction perpendicular to the field and the gravitational potential.
    \item We find that the total mass of infalling cold gas associated with the original cloud increases with time due to condensation of mixed gas despite the low metallicity of the cloud. This is in agreement with \cite{gritton17} and analogous to the trend seen in simulations of higher metallicity galactic fountain clouds \citep{marinacci10,armillotta16,gronnow18}. As found for the clouds in \cite{gronnow18} the role of the magnetic field is to partially suppress the efficiency of the condensation by limiting stripping and mixing.
    \item For our strong magnetic field case, magnetic tension leads to severe deceleration of mostly condensed cloudlets which leads to a bimodal velocity distribution of cold gas with a low-velocity bump at $|v_z|<60 \kms$. These cloudlets are generally small so due to their slow infall velocities they might disperse before being accreted onto the disc. The remnants of the original cloud which represents the majority of the cold gas is however not significantly decelerated.
\end{enumerate}
Overall, HVCs formed through thermal instability in the corona appear to be able to reach the disc and feed star formation. Additionally, condensation along their trajectories may moderately add to this accretion. However, ignoring the magnetic field substantially overestimates the amount of gas that may be accreted in this way.

\section*{Acknowledgements}
We thank the referee for a constructive report that improved the quality of the paper.

AG and FF acknowledge support from the Netherlands Research School for Astronomy (Nederlandse Onderzoekschool voor Astronomie, NOVA), Network 1, Project 10.1.5.7.

AG and TTG acknowledge financial support from the Australian Research Council (ARC) through an Australian Laureate Fellowship awarded to JBH. We acknowledge the facilities, and the scientific and technical assistance of the Sydney Informatics Hub at the University of Sydney and, in particular, access to the high-performance computing facility Artemis and additional resources on the National Computational Infrastructure (NCI) through the University of Sydney's Grand Challenge Program `Astrophysics Grand Challenge: From Large to Small' (CIs G. F. Lewis and J. Bland-Hawthorn). We also acknowledge additional access to NCI facilities through the Astronomy Supercomputer Time Allocation Committee (ASTAC) scheme managed by Astronomy Australia Limited and supported by the Australian Government.
AG and FF would like to thank the Center for Information Technology of the University of Groningen for their support and for providing access to the Peregrine high performance computing cluster.

\section*{Data availability}
The code that we used to generate our simulations is available at \url{https://github.com/agronnow/hvc_gravity}. Simulation output data will be shared on reasonable request to the corresponding author.

\bibliographystyle{mnras}
\bibliography{AG_hvcsurvival}

\begin{thebibliography}{}
\makeatletter
\relax
\def\mn@urlcharsother{\let\do\@makeother \do\$\do\&\do\#\do\^\do\_\do\%\do\~}
\def\mn@doi{\begingroup\mn@urlcharsother \@ifnextchar [ {\mn@doi@}
  {\mn@doi@[]}}
\def\mn@doi@[#1]#2{\def\@tempa{#1}\ifx\@tempa\@empty \href
  {http://dx.doi.org/#2} {doi:#2}\else \href {http://dx.doi.org/#2} {#1}\fi
  \endgroup}
\def\mn@eprint#1#2{\mn@eprint@#1:#2::\@nil}
\def\mn@eprint@arXiv#1{\href {http://arxiv.org/abs/#1} {{\tt arXiv:#1}}}
\def\mn@eprint@dblp#1{\href {http://dblp.uni-trier.de/rec/bibtex/#1.xml}
  {dblp:#1}}
\def\mn@eprint@#1:#2:#3:#4\@nil{\def\@tempa {#1}\def\@tempb {#2}\def\@tempc
  {#3}\ifx \@tempc \@empty \let \@tempc \@tempb \let \@tempb \@tempa \fi \ifx
  \@tempb \@empty \def\@tempb {arXiv}\fi \@ifundefined
  {mn@eprint@\@tempb}{\@tempb:\@tempc}{\expandafter \expandafter \csname
  mn@eprint@\@tempb\endcsname \expandafter{\@tempc}}}

\bibitem[\protect\citeauthoryear{{Armillotta}, {Fraternali}  \&
  {Marinacci}}{{Armillotta} et~al.}{2016}]{armillotta16}
{Armillotta} L.,  {Fraternali} F.,   {Marinacci} F.,  2016, \mn@doi [\mnras]
  {10.1093/mnras/stw1930}, \href
  {http://adsabs.harvard.edu/abs/2016MNRAS.462.4157A} {462, 4157}

\bibitem[\protect\citeauthoryear{{Armillotta}, {Fraternali}, {Werk},
  {Prochaska}  \& {Marinacci}}{{Armillotta} et~al.}{2017}]{armillotta17}
{Armillotta} L.,  {Fraternali} F.,  {Werk} J.~K.,  {Prochaska} J.~X.,
  {Marinacci} F.,  2017, \mn@doi [\mnras] {10.1093/mnras/stx1239}, \href
  {http://adsabs.harvard.edu/abs/2017MNRAS.470..114A} {470, 114}

\bibitem[\protect\citeauthoryear{{Asai}, {Fukuda}  \& {Matsumoto}}{{Asai}
  et~al.}{2007}]{asai07}
{Asai} N.,  {Fukuda} N.,   {Matsumoto} R.,  2007, \mn@doi [\apj]
  {10.1086/518235}, \href
  {https://ui.adsabs.harvard.edu/abs/2007ApJ...663..816A} {663, 816}

\bibitem[\protect\citeauthoryear{{Banda-Barrag{\'a}n}, {Parkin}, {Federrath},
  {Crocker}  \& {Bicknell}}{{Banda-Barrag{\'a}n}
  et~al.}{2016}]{banda-barragan16}
{Banda-Barrag{\'a}n} W.~E.,  {Parkin} E.~R.,  {Federrath} C.,  {Crocker} R.~M.,
    {Bicknell} G.~V.,  2016, \mn@doi [\mnras] {10.1093/mnras/stv2405}, \href
  {http://adsabs.harvard.edu/abs/2016MNRAS.455.1309B} {455, 1309}

\bibitem[\protect\citeauthoryear{{Banda-Barrag{\'a}n}, {Federrath}, {Crocker}
  \& {Bicknell}}{{Banda-Barrag{\'a}n} et~al.}{2018}]{banda-barragan18}
{Banda-Barrag{\'a}n} W.~E.,  {Federrath} C.,  {Crocker} R.~M.,   {Bicknell}
  G.~V.,  2018, \mn@doi [\mnras] {10.1093/mnras/stx2541}, \href
  {http://adsabs.harvard.edu/abs/2018MNRAS.473.3454B} {473, 3454}

\bibitem[\protect\citeauthoryear{{Barger} et~al.,}{{Barger}
  et~al.}{2020}]{barger20}
{Barger} K.~A.,  et~al., 2020, \mn@doi [\apj] {10.3847/1538-4357/abb376}, \href
  {https://ui.adsabs.harvard.edu/abs/2020ApJ...902..154B} {902, 154}

\bibitem[\protect\citeauthoryear{{Beck}}{{Beck}}{2016}]{beck16}
{Beck} R.,  2016, \mn@doi [\aapr] {10.1007/s00159-015-0084-4}, \href
  {http://adsabs.harvard.edu/abs/2016A%26ARv..24....4B} {24, 4}

\bibitem[\protect\citeauthoryear{{Berry}}{{Berry}}{2015}]{berry15}
{Berry} D.~S.,  2015, \mn@doi [Astronomy and Computing]
  {10.1016/j.ascom.2014.11.004}, \href
  {https://ui.adsabs.harvard.edu/abs/2015A&C....10...22B} {10, 22}

\bibitem[\protect\citeauthoryear{{Berry}, {Reinhold}, {Jenness}  \&
  {Economou}}{{Berry} et~al.}{2007}]{berry07}
{Berry} D.~S.,  {Reinhold} K.,  {Jenness} T.,   {Economou} F.,  2007, in {Shaw}
  R.~A.,  {Hill} F.,   {Bell} D.~J.,  eds,  Astronomical Society of the Pacific
  Conference Series Vol. 376, Astronomical Data Analysis Software and Systems
  XVI. p.~425

\bibitem[\protect\citeauthoryear{{Binney}, {Nipoti}  \& {Fraternali}}{{Binney}
  et~al.}{2009}]{binney09}
{Binney} J.,  {Nipoti} C.,   {Fraternali} F.,  2009, \mn@doi [\mnras]
  {10.1111/j.1365-2966.2009.15113.x}, \href
  {http://adsabs.harvard.edu/abs/2009MNRAS.397.1804B} {397, 1804}

\bibitem[\protect\citeauthoryear{{Bish}, {Werk}, {Prochaska}, {Rubin}, {Zheng},
  {O'Meara}  \& {Deason}}{{Bish} et~al.}{2019}]{bish19}
{Bish} H.~V.,  {Werk} J.~K.,  {Prochaska} J.~X.,  {Rubin} K. H.~R.,  {Zheng}
  Y.,  {O'Meara} J.~M.,   {Deason} A.~J.,  2019, \mn@doi [\apj]
  {10.3847/1538-4357/ab3414}, \href
  {https://ui.adsabs.harvard.edu/abs/2019ApJ...882...76B} {882, 76}

\bibitem[\protect\citeauthoryear{{Bland-Hawthorn}, {Sutherland}, {Agertz}  \&
  {Moore}}{{Bland-Hawthorn} et~al.}{2007}]{bland-hawthorn07}
{Bland-Hawthorn} J.,  {Sutherland} R.,  {Agertz} O.,   {Moore} B.,  2007,
  \mn@doi [\apjl] {10.1086/524657}, \href
  {http://adsabs.harvard.edu/abs/2007ApJ...670L.109B} {670, L109}

\bibitem[\protect\citeauthoryear{{Bovy}}{{Bovy}}{2015}]{bovy15}
{Bovy} J.,  2015, \mn@doi [\apjs] {10.1088/0067-0049/216/2/29}, \href
  {https://ui.adsabs.harvard.edu/abs/2015ApJS..216...29B} {216, 29}

\bibitem[\protect\citeauthoryear{{Bregman}}{{Bregman}}{2007}]{bregman07}
{Bregman} J.~N.,  2007, \mn@doi [\araa]
  {10.1146/annurev.astro.45.051806.110619}, \href
  {http://adsabs.harvard.edu/abs/2007ARA%26A..45..221B} {45, 221}

\bibitem[\protect\citeauthoryear{{Bregman}, {Anderson}, {Miller},
  {Hodges-Kluck}, {Dai}, {Li}, {Li}  \& {Qu}}{{Bregman}
  et~al.}{2018}]{bregman18}
{Bregman} J.~N.,  {Anderson} M.~E.,  {Miller} M.~J.,  {Hodges-Kluck} E.,  {Dai}
  X.,  {Li} J.-T.,  {Li} Y.,   {Qu} Z.,  2018, \mn@doi [\apj]
  {10.3847/1538-4357/aacafe}, \href
  {https://ui.adsabs.harvard.edu/abs/2018ApJ...862....3B} {862, 3}

\bibitem[\protect\citeauthoryear{{Cautun} et~al.,}{{Cautun}
  et~al.}{2020}]{cautun20}
{Cautun} M.,  et~al., 2020, \mn@doi [\mnras] {10.1093/mnras/staa1017}, \href
  {https://ui.adsabs.harvard.edu/abs/2020MNRAS.494.4291C} {494, 4291}

\bibitem[\protect\citeauthoryear{{Cottle}, {Scannapieco}, {Br{\"u}ggen},
  {Banda-Barrag{\'a}n}  \& {Federrath}}{{Cottle} et~al.}{2020}]{cottle20}
{Cottle} J.,  {Scannapieco} E.,  {Br{\"u}ggen} M.,  {Banda-Barrag{\'a}n} W.,
  {Federrath} C.,  2020, \mn@doi [\apj] {10.3847/1538-4357/ab76d1}, \href
  {https://ui.adsabs.harvard.edu/abs/2020ApJ...892...59C} {892, 59}

\bibitem[\protect\citeauthoryear{{Dursi} \& {Pfrommer}}{{Dursi} \&
  {Pfrommer}}{2008}]{dursi08}
{Dursi} L.~J.,  {Pfrommer} C.,  2008, \mn@doi [\apj] {10.1086/529371}, \href
  {http://adsabs.harvard.edu/abs/2008ApJ...677..993D} {677, 993}

\bibitem[\protect\citeauthoryear{{Dutta} \& {Sharma}}{{Dutta} \&
  {Sharma}}{2019}]{dutta19}
{Dutta} A.,  {Sharma} P.,  2019, \mn@doi [Research Notes of the American
  Astronomical Society] {10.3847/2515-5172/ab4bd8}, \href
  {https://ui.adsabs.harvard.edu/abs/2019RNAAS...3..148D} {3, 148}

\bibitem[\protect\citeauthoryear{{Edmunds}}{{Edmunds}}{1990}]{edmunds90}
{Edmunds} M.~G.,  1990, \mnras, \href
  {https://ui.adsabs.harvard.edu/abs/1990MNRAS.246..678E} {246, 678}

\bibitem[\protect\citeauthoryear{{Faerman}, {Sternberg}  \& {McKee}}{{Faerman}
  et~al.}{2017}]{faerman17}
{Faerman} Y.,  {Sternberg} A.,   {McKee} C.~F.,  2017, \mn@doi [\apj]
  {10.3847/1538-4357/835/1/52}, \href
  {http://adsabs.harvard.edu/abs/2017ApJ...835...52F} {835, 52}

\bibitem[\protect\citeauthoryear{{Ferland}, {Korista}, {Verner}, {Ferguson},
  {Kingdon}  \& {Verner}}{{Ferland} et~al.}{1998}]{ferland98}
{Ferland} G.~J.,  {Korista} K.~T.,  {Verner} D.~A.,  {Ferguson} J.~W.,
  {Kingdon} J.~B.,   {Verner} E.~M.,  1998, \mn@doi [\pasp] {10.1086/316190},
  \href {http://adsabs.harvard.edu/abs/1998PASP..110..761F} {110, 761}

\bibitem[\protect\citeauthoryear{{Fox} et~al.,}{{Fox} et~al.}{2016}]{fox16}
{Fox} A.~J.,  et~al., 2016, \mn@doi [\apjl] {10.3847/2041-8205/816/1/L11},
  \href {http://adsabs.harvard.edu/abs/2016ApJ...816L..11F} {816, L11}

\bibitem[\protect\citeauthoryear{{Fox}, {Richter}, {Ashley}, {Heckman},
  {Lehner}, {Werk}, {Bordoloi}  \& {Peeples}}{{Fox} et~al.}{2019}]{fox19}
{Fox} A.~J.,  {Richter} P.,  {Ashley} T.,  {Heckman} T.~M.,  {Lehner} N.,
  {Werk} J.~K.,  {Bordoloi} R.,   {Peeples} M.~S.,  2019, \mn@doi [\apj]
  {10.3847/1538-4357/ab40ad}, \href
  {https://ui.adsabs.harvard.edu/abs/2019ApJ...884...53F} {884, 53}

\bibitem[\protect\citeauthoryear{{Fraternali}, {Marasco}, {Armillotta}  \&
  {Marinacci}}{{Fraternali} et~al.}{2015}]{fraternali15}
{Fraternali} F.,  {Marasco} A.,  {Armillotta} L.,   {Marinacci} F.,  2015,
  \mn@doi [\mnras] {10.1093/mnrasl/slu182}, \href
  {http://adsabs.harvard.edu/abs/2015MNRAS.447L..70F} {447, L70}

\bibitem[\protect\citeauthoryear{{Fromang}, {Hennebelle}  \&
  {Teyssier}}{{Fromang} et~al.}{2006}]{fromang06}
{Fromang} S.,  {Hennebelle} P.,   {Teyssier} R.,  2006, \mn@doi [\aap]
  {10.1051/0004-6361:20065371}, \href
  {https://ui.adsabs.harvard.edu/abs/2006A&A...457..371F} {457, 371}

\bibitem[\protect\citeauthoryear{{Gaensler}, {Madsen}, {Chatterjee}  \&
  {Mao}}{{Gaensler} et~al.}{2008}]{gaensler08}
{Gaensler} B.~M.,  {Madsen} G.~J.,  {Chatterjee} S.,   {Mao} S.~A.,  2008,
  \mn@doi [\pasa] {10.1071/AS08004}, \href
  {https://ui.adsabs.harvard.edu/abs/2008PASA...25..184G} {25, 184}

\bibitem[\protect\citeauthoryear{{Galyardt} \& {Shelton}}{{Galyardt} \&
  {Shelton}}{2016}]{galyardt16}
{Galyardt} J.,  {Shelton} R.~L.,  2016, \mn@doi [\apjl]
  {10.3847/2041-8205/816/1/L18}, \href
  {http://adsabs.harvard.edu/abs/2016ApJ...816L..18G} {816, L18}

\bibitem[\protect\citeauthoryear{{Grcevich} \& {Putman}}{{Grcevich} \&
  {Putman}}{2009}]{grcevich09}
{Grcevich} J.,  {Putman} M.~E.,  2009, \mn@doi [\apj]
  {10.1088/0004-637X/696/1/385}, \href
  {https://ui.adsabs.harvard.edu/abs/2009ApJ...696..385G} {696, 385}

\bibitem[\protect\citeauthoryear{{Gregori}, {Miniati}, {Ryu}  \&
  {Jones}}{{Gregori} et~al.}{1999}]{gregori99}
{Gregori} G.,  {Miniati} F.,  {Ryu} D.,   {Jones} T.~W.,  1999, \mn@doi [\apjl]
  {10.1086/312402}, \href {http://adsabs.harvard.edu/abs/1999ApJ...527L.113G}
  {527, L113}

\bibitem[\protect\citeauthoryear{{Gregori}, {Miniati}, {Ryu}  \&
  {Jones}}{{Gregori} et~al.}{2000}]{gregori00}
{Gregori} G.,  {Miniati} F.,  {Ryu} D.,   {Jones} T.~W.,  2000, \mn@doi [\apj]
  {10.1086/317130}, \href {http://adsabs.harvard.edu/abs/2000ApJ...543..775G}
  {543, 775}

\bibitem[\protect\citeauthoryear{{Gritton}, {Shelton}  \& {Galyardt}}{{Gritton}
  et~al.}{2017}]{gritton17}
{Gritton} J.~A.,  {Shelton} R.~L.,   {Galyardt} J.~E.,  2017, \mn@doi [\apj]
  {10.3847/1538-4357/aa756d}, \href
  {http://adsabs.harvard.edu/abs/2017ApJ...842..102G} {842, 102}

\bibitem[\protect\citeauthoryear{{Gronke} \& {Oh}}{{Gronke} \&
  {Oh}}{2020a}]{gronke20a}
{Gronke} M.,  {Oh} S.~P.,  2020a, \mn@doi [\mnras] {10.1093/mnras/stz3332},
  \href {https://ui.adsabs.harvard.edu/abs/2020MNRAS.492.1970G} {492, 1970}

\bibitem[\protect\citeauthoryear{{Gronke} \& {Oh}}{{Gronke} \&
  {Oh}}{2020b}]{gronke20b}
{Gronke} M.,  {Oh} S.~P.,  2020b, \mn@doi [\mnras] {10.1093/mnrasl/slaa033},
  \href {https://ui.adsabs.harvard.edu/abs/2020MNRAS.494L..27G} {494, L27}

\bibitem[\protect\citeauthoryear{{Gr{\o}nnow}, {Tepper-Garc{\'{\i}}a},
  {Bland-Hawthorn}  \& {McClure-Griffiths}}{{Gr{\o}nnow}
  et~al.}{2017}]{gronnow17}
{Gr{\o}nnow} A.,  {Tepper-Garc{\'{\i}}a} T.,  {Bland-Hawthorn} J.,
  {McClure-Griffiths} N.~M.,  2017, \mn@doi [\apj] {10.3847/1538-4357/aa7ed2},
  \href {http://adsabs.harvard.edu/abs/2017ApJ...845...69G} {845, 69}

\bibitem[\protect\citeauthoryear{{Gr{\o}nnow}, {Tepper-Garc{\'\i}a}  \&
  {Bland-Hawthorn}}{{Gr{\o}nnow} et~al.}{2018}]{gronnow18}
{Gr{\o}nnow} A.,  {Tepper-Garc{\'\i}a} T.,   {Bland-Hawthorn} J.,  2018,
  \mn@doi [\apj] {10.3847/1538-4357/aada0e}, \href
  {https://ui.adsabs.harvard.edu/abs/2018ApJ...865...64G} {865, 64}

\bibitem[\protect\citeauthoryear{{Haffner}, {Reynolds}, {Tufte}, {Madsen},
  {Jaehnig}  \& {Percival}}{{Haffner} et~al.}{2003}]{haffner03}
{Haffner} L.~M.,  {Reynolds} R.~J.,  {Tufte} S.~L.,  {Madsen} G.~J.,  {Jaehnig}
  K.~P.,   {Percival} J.~W.,  2003, \mn@doi [\apjs] {10.1086/378850}, \href
  {https://ui.adsabs.harvard.edu/abs/2003ApJS..149..405H} {149, 405}

\bibitem[\protect\citeauthoryear{{Heitsch} \& {Putman}}{{Heitsch} \&
  {Putman}}{2009}]{heitsch09}
{Heitsch} F.,  {Putman} M.~E.,  2009, \mn@doi [\apj]
  {10.1088/0004-637X/698/2/1485}, \href
  {http://adsabs.harvard.edu/abs/2009ApJ...698.1485H} {698, 1485}

\bibitem[\protect\citeauthoryear{{Heitsch}, {Marchal}, {Miville-Desch{\^e}nes},
  {Shull}  \& {Fox}}{{Heitsch} et~al.}{2021}]{heitsch21}
{Heitsch} F.,  {Marchal} A.,  {Miville-Desch{\^e}nes} M.~A.,  {Shull} J.~M.,
  {Fox} A.~J.,  2021, \mn@doi [\mnras] {10.1093/mnras/stab3266}, \href
  {https://ui.adsabs.harvard.edu/abs/2021MNRAS.tmp.3006H} {}

\bibitem[\protect\citeauthoryear{{Henley} \& {Shelton}}{{Henley} \&
  {Shelton}}{2015}]{henley15}
{Henley} D.~B.,  {Shelton} R.~L.,  2015, \mn@doi [\apj]
  {10.1088/0004-637X/808/1/22}, \href
  {http://adsabs.harvard.edu/abs/2015ApJ...808...22H} {808, 22}

\bibitem[\protect\citeauthoryear{{Hodges-Kluck}, {Miller}  \&
  {Bregman}}{{Hodges-Kluck} et~al.}{2016}]{hodges-kluck16}
{Hodges-Kluck} E.~J.,  {Miller} M.~J.,   {Bregman} J.~N.,  2016, \mn@doi [\apj]
  {10.3847/0004-637X/822/1/21}, \href
  {https://ui.adsabs.harvard.edu/abs/2016ApJ...822...21H} {822, 21}

\bibitem[\protect\citeauthoryear{{Hodges-Kluck}, {Bregman}  \&
  {Li}}{{Hodges-Kluck} et~al.}{2018}]{hodges-kluck18}
{Hodges-Kluck} E.~J.,  {Bregman} J.~N.,   {Li} J.-t.,  2018, \mn@doi [\apj]
  {10.3847/1538-4357/aae38a}, \href
  {https://ui.adsabs.harvard.edu/abs/2018ApJ...866..126H} {866, 126}

\bibitem[\protect\citeauthoryear{{Jansson} \& {Farrar}}{{Jansson} \&
  {Farrar}}{2012a}]{jansson12a}
{Jansson} R.,  {Farrar} G.~R.,  2012a, \mn@doi [\apj]
  {10.1088/0004-637X/757/1/14}, \href
  {http://adsabs.harvard.edu/abs/2012ApJ...757...14J} {757, 14}

\bibitem[\protect\citeauthoryear{{Jansson} \& {Farrar}}{{Jansson} \&
  {Farrar}}{2012b}]{jansson12b}
{Jansson} R.,  {Farrar} G.~R.,  2012b, \mn@doi [\apjl]
  {10.1088/2041-8205/761/1/L11}, \href
  {http://adsabs.harvard.edu/abs/2012ApJ...761L..11J} {761, L11}

\bibitem[\protect\citeauthoryear{{Ji}, {Peng Oh}  \& {McCourt}}{{Ji}
  et~al.}{2018}]{ji18}
{Ji} S.,  {Peng Oh} S.,   {McCourt} M.,  2018, \mn@doi [\mnras]
  {10.1093/mnras/sty293}, \href
  {http://adsabs.harvard.edu/abs/2018MNRAS.tmp..287J} {}

\bibitem[\protect\citeauthoryear{{Jones}, {Ryu}  \& {Tregillis}}{{Jones}
  et~al.}{1996}]{jones96}
{Jones} T.~W.,  {Ryu} D.,   {Tregillis} I.~L.,  1996, \mn@doi [\apj]
  {10.1086/178151}, \href {http://adsabs.harvard.edu/abs/1996ApJ...473..365J}
  {473, 365}

\bibitem[\protect\citeauthoryear{{Joung}, {Bryan}  \& {Putman}}{{Joung}
  et~al.}{2012}]{joung12}
{Joung} M.~R.,  {Bryan} G.~L.,   {Putman} M.~E.,  2012, \mn@doi [\apj]
  {10.1088/0004-637X/745/2/148}, \href
  {http://adsabs.harvard.edu/abs/2012ApJ...745..148J} {745, 148}

\bibitem[\protect\citeauthoryear{{Kalberla} \& {Dedes}}{{Kalberla} \&
  {Dedes}}{2008}]{kalberla08}
{Kalberla} P.~M.~W.,  {Dedes} L.,  2008, \mn@doi [\aap]
  {10.1051/0004-6361:20079240}, \href
  {https://ui.adsabs.harvard.edu/abs/2008A&A...487..951K} {487, 951}

\bibitem[\protect\citeauthoryear{{Koch} et~al.,}{{Koch} et~al.}{2018}]{koch18}
{Koch} E.~W.,  et~al., 2018, \mn@doi [\mnras] {10.1093/mnras/sty1674}, \href
  {https://ui.adsabs.harvard.edu/abs/2018MNRAS.479.2505K} {479, 2505}

\bibitem[\protect\citeauthoryear{{Kooij}, {Gr{\o}nnow}  \&
  {Fraternali}}{{Kooij} et~al.}{2021}]{kooij21}
{Kooij} R.,  {Gr{\o}nnow} A.,   {Fraternali} F.,  2021, \mn@doi [\mnras]
  {10.1093/mnras/stab110}, \href
  {https://ui.adsabs.harvard.edu/abs/2021MNRAS.502.1263K} {502, 1263}

\bibitem[\protect\citeauthoryear{{Kwak}, {Shelton}  \& {Raley}}{{Kwak}
  et~al.}{2009}]{kwak09}
{Kwak} K.,  {Shelton} R.~L.,   {Raley} E.~A.,  2009, \mn@doi [\apj]
  {10.1088/0004-637X/699/2/1775}, \href
  {https://ui.adsabs.harvard.edu/abs/2009ApJ...699.1775K} {699, 1775}

\bibitem[\protect\citeauthoryear{{Larson}}{{Larson}}{1972}]{larson72}
{Larson} R.~B.,  1972, \mn@doi [\nat] {10.1038/236021a0}, \href
  {https://ui.adsabs.harvard.edu/abs/1972Natur.236...21L} {236, 21}

\bibitem[\protect\citeauthoryear{{Lehner} \& {Howk}}{{Lehner} \&
  {Howk}}{2010}]{lehner10}
{Lehner} N.,  {Howk} J.~C.,  2010, \mn@doi [\apjl]
  {10.1088/2041-8205/709/2/L138}, \href
  {https://ui.adsabs.harvard.edu/abs/2010ApJ...709L.138L} {709, L138}

\bibitem[\protect\citeauthoryear{{Lehner}, {Howk}, {Thom}, {Fox}, {Tumlinson},
  {Tripp}  \& {Meiring}}{{Lehner} et~al.}{2012}]{lehner12}
{Lehner} N.,  {Howk} J.~C.,  {Thom} C.,  {Fox} A.~J.,  {Tumlinson} J.,  {Tripp}
  T.~M.,   {Meiring} J.~D.,  2012, \mn@doi [\mnras]
  {10.1111/j.1365-2966.2012.21428.x}, \href
  {http://adsabs.harvard.edu/abs/2012MNRAS.424.2896L} {424, 2896}

\bibitem[\protect\citeauthoryear{{Li}, {Hopkins}, {Squire}  \& {Hummels}}{{Li}
  et~al.}{2020}]{li20}
{Li} Z.,  {Hopkins} P.~F.,  {Squire} J.,   {Hummels} C.,  2020, \mn@doi
  [\mnras] {10.1093/mnras/stz3567}, \href
  {https://ui.adsabs.harvard.edu/abs/2020MNRAS.492.1841L} {492, 1841}

\bibitem[\protect\citeauthoryear{{Licquia} \& {Newman}}{{Licquia} \&
  {Newman}}{2015}]{licquia15}
{Licquia} T.~C.,  {Newman} J.~A.,  2015, \mn@doi [\apj]
  {10.1088/0004-637X/806/1/96}, \href
  {https://ui.adsabs.harvard.edu/abs/2015ApJ...806...96L} {806, 96}

\bibitem[\protect\citeauthoryear{{Maller} \& {Bullock}}{{Maller} \&
  {Bullock}}{2004}]{maller04}
{Maller} A.~H.,  {Bullock} J.~S.,  2004, \mn@doi [\mnras]
  {10.1111/j.1365-2966.2004.08349.x}, \href
  {https://ui.adsabs.harvard.edu/abs/2004MNRAS.355..694M} {355, 694}

\bibitem[\protect\citeauthoryear{{Marasco} \& {Fraternali}}{{Marasco} \&
  {Fraternali}}{2017}]{marasco17}
{Marasco} A.,  {Fraternali} F.,  2017, \mn@doi [\mnras]
  {10.1093/mnrasl/slw195}, \href
  {http://adsabs.harvard.edu/abs/2017MNRAS.464L.100M} {464, L100}

\bibitem[\protect\citeauthoryear{{Marinacci}, {Binney}, {Fraternali}, {Nipoti},
  {Ciotti}  \& {Londrillo}}{{Marinacci} et~al.}{2010}]{marinacci10}
{Marinacci} F.,  {Binney} J.,  {Fraternali} F.,  {Nipoti} C.,  {Ciotti} L.,
  {Londrillo} P.,  2010, \mn@doi [\mnras] {10.1111/j.1365-2966.2010.16352.x},
  \href {http://adsabs.harvard.edu/abs/2010MNRAS.404.1464M} {404, 1464}

\bibitem[\protect\citeauthoryear{{McCourt}, {O'Leary}, {Madigan}  \&
  {Quataert}}{{McCourt} et~al.}{2015}]{mccourt15}
{McCourt} M.,  {O'Leary} R.~M.,  {Madigan} A.-M.,   {Quataert} E.,  2015,
  \mn@doi [\mnras] {10.1093/mnras/stv355}, \href
  {http://adsabs.harvard.edu/abs/2015MNRAS.449....2M} {449, 2}

\bibitem[\protect\citeauthoryear{{McCourt}, {Oh}, {O'Leary}  \&
  {Madigan}}{{McCourt} et~al.}{2018}]{mccourt18}
{McCourt} M.,  {Oh} S.~P.,  {O'Leary} R.,   {Madigan} A.-M.,  2018, \mn@doi
  [\mnras] {10.1093/mnras/stx2687}, \href
  {https://ui.adsabs.harvard.edu/abs/2018MNRAS.473.5407M} {473, 5407}

\bibitem[\protect\citeauthoryear{{Miller} \& {Bregman}}{{Miller} \&
  {Bregman}}{2015}]{miller15}
{Miller} M.~J.,  {Bregman} J.~N.,  2015, \mn@doi [\apj]
  {10.1088/0004-637X/800/1/14}, \href
  {http://adsabs.harvard.edu/abs/2015ApJ...800...14M} {800, 14}

\bibitem[\protect\citeauthoryear{{Muller}, {Oort}  \& {Raimond}}{{Muller}
  et~al.}{1963}]{muller63}
{Muller} C.~A.,  {Oort} J.~H.,   {Raimond} E.,  1963, Academie des Sciences
  Paris Comptes Rendus, \href
  {http://adsabs.harvard.edu/abs/1963CRAS..257.1661M} {257, 1661}

\bibitem[\protect\citeauthoryear{{Nichols} \& {Bland-Hawthorn}}{{Nichols} \&
  {Bland-Hawthorn}}{2009}]{nichols09}
{Nichols} M.,  {Bland-Hawthorn} J.,  2009, \mn@doi [\apj]
  {10.1088/0004-637X/707/2/1642}, \href
  {http://adsabs.harvard.edu/abs/2009ApJ...707.1642N} {707, 1642}

\bibitem[\protect\citeauthoryear{{Nichols}, {Mirabal}, {Agertz}, {Lockman}  \&
  {Bland-Hawthorn}}{{Nichols} et~al.}{2014}]{nichols14}
{Nichols} M.,  {Mirabal} N.,  {Agertz} O.,  {Lockman} F.~J.,   {Bland-Hawthorn}
  J.,  2014, \mn@doi [\mnras] {10.1093/mnras/stu1028}, \href
  {http://adsabs.harvard.edu/abs/2014MNRAS.442.2883N} {442, 2883}

\bibitem[\protect\citeauthoryear{{Peek}, {Bordoloi}, {Sana}, {Roman-Duval},
  {Tumlinson}  \& {Zheng}}{{Peek} et~al.}{2016}]{peek16}
{Peek} J.~E.~G.,  {Bordoloi} R.,  {Sana} H.,  {Roman-Duval} J.,  {Tumlinson}
  J.,   {Zheng} Y.,  2016, \mn@doi [\apjl] {10.3847/2041-8205/828/2/L20}, \href
  {https://ui.adsabs.harvard.edu/abs/2016ApJ...828L..20P} {828, L20}

\bibitem[\protect\citeauthoryear{{Pezzulli} \& {Fraternali}}{{Pezzulli} \&
  {Fraternali}}{2016}]{pezzulli16}
{Pezzulli} G.,  {Fraternali} F.,  2016, \mn@doi [\mnras]
  {10.1093/mnras/stv2397}, \href
  {https://ui.adsabs.harvard.edu/abs/2016MNRAS.455.2308P} {455, 2308}

\bibitem[\protect\citeauthoryear{{Pezzulli}, {Fraternali}  \&
  {Binney}}{{Pezzulli} et~al.}{2017}]{pezzulli17}
{Pezzulli} G.,  {Fraternali} F.,   {Binney} J.,  2017, \mn@doi [\mnras]
  {10.1093/mnras/stx029}, \href
  {https://ui.adsabs.harvard.edu/abs/2017MNRAS.467..311P} {467, 311}

\bibitem[\protect\citeauthoryear{{Posti} \& {Helmi}}{{Posti} \&
  {Helmi}}{2019}]{posti19}
{Posti} L.,  {Helmi} A.,  2019, \mn@doi [\aap] {10.1051/0004-6361/201833355},
  \href {https://ui.adsabs.harvard.edu/abs/2019A&A...621A..56P} {621, A56}

\bibitem[\protect\citeauthoryear{{Putman}, {Bland-Hawthorn}, {Veilleux},
  {Gibson}, {Freeman}  \& {Maloney}}{{Putman} et~al.}{2003}]{putman03b}
{Putman} M.~E.,  {Bland-Hawthorn} J.,  {Veilleux} S.,  {Gibson} B.~K.,
  {Freeman} K.~C.,   {Maloney} P.~R.,  2003, \mn@doi [\apj] {10.1086/378555},
  \href {https://ui.adsabs.harvard.edu/abs/2003ApJ...597..948P} {597, 948}

\bibitem[\protect\citeauthoryear{{Putman}, {Saul}  \& {Mets}}{{Putman}
  et~al.}{2011}]{putman11}
{Putman} M.~E.,  {Saul} D.~R.,   {Mets} E.,  2011, \mn@doi [\mnras]
  {10.1111/j.1365-2966.2011.19524.x}, \href
  {http://adsabs.harvard.edu/abs/2011MNRAS.418.1575P} {418, 1575}

\bibitem[\protect\citeauthoryear{{Putman}, {Peek}  \& {Joung}}{{Putman}
  et~al.}{2012}]{putman12}
{Putman} M.~E.,  {Peek} J.~E.~G.,   {Joung} M.~R.,  2012, \mn@doi [\araa]
  {10.1146/annurev-astro-081811-125612}, \href
  {http://adsabs.harvard.edu/abs/2012ARA%26A..50..491P} {50, 491}

\bibitem[\protect\citeauthoryear{{Richter}, {de Boer}, {Werner}  \&
  {Rauch}}{{Richter} et~al.}{2015}]{richter15}
{Richter} P.,  {de Boer} K.~S.,  {Werner} K.,   {Rauch} T.,  2015, \mn@doi
  [\aap] {10.1051/0004-6361/201527451}, \href
  {https://ui.adsabs.harvard.edu/abs/2015A&A...584L...6R} {584, L6}

\bibitem[\protect\citeauthoryear{{Robitaille} \& {Whitney}}{{Robitaille} \&
  {Whitney}}{2010}]{robitaille10}
{Robitaille} T.~P.,  {Whitney} B.~A.,  2010, \mn@doi [\apjl]
  {10.1088/2041-8205/710/1/L11}, \href
  {http://adsabs.harvard.edu/abs/2010ApJ...710L..11R} {710, L11}

\bibitem[\protect\citeauthoryear{{Sander} \& {Hensler}}{{Sander} \&
  {Hensler}}{2019}]{sander19}
{Sander} B.,  {Hensler} G.,  2019, \mn@doi [\mnras] {10.1093/mnrasl/slz144},
  \href {https://ui.adsabs.harvard.edu/abs/2019MNRAS.490L..52S} {490, L52}

\bibitem[\protect\citeauthoryear{{Sander} \& {Hensler}}{{Sander} \&
  {Hensler}}{2021}]{sander21}
{Sander} B.,  {Hensler} G.,  2021, \mn@doi [\mnras] {10.1093/mnras/staa3952},
  \href {https://ui.adsabs.harvard.edu/abs/2021MNRAS.501.5330S} {501, 5330}

\bibitem[\protect\citeauthoryear{{Santill{\'a}n}, {Franco}, {Martos}  \&
  {Kim}}{{Santill{\'a}n} et~al.}{1999}]{santillan99}
{Santill{\'a}n} A.,  {Franco} J.,  {Martos} M.,   {Kim} J.,  1999, \mn@doi
  [\apj] {10.1086/307065}, \href
  {https://ui.adsabs.harvard.edu/abs/1999ApJ...515..657S} {515, 657}

\bibitem[\protect\citeauthoryear{{Scannapieco} \& {Br{\"u}ggen}}{{Scannapieco}
  \& {Br{\"u}ggen}}{2015}]{scannapieco15}
{Scannapieco} E.,  {Br{\"u}ggen} M.,  2015, \mn@doi [\apj]
  {10.1088/0004-637X/805/2/158}, \href
  {http://adsabs.harvard.edu/abs/2015ApJ...805..158S} {805, 158}

\bibitem[\protect\citeauthoryear{{Schlesinger} et~al.,}{{Schlesinger}
  et~al.}{2012}]{schlesinger12}
{Schlesinger} K.~J.,  et~al., 2012, \mn@doi [\apj]
  {10.1088/0004-637X/761/2/160}, \href
  {http://adsabs.harvard.edu/abs/2012ApJ...761..160S} {761, 160}

\bibitem[\protect\citeauthoryear{{Sch{\"o}nrich} \& {McMillan}}{{Sch{\"o}nrich}
  \& {McMillan}}{2017}]{schoenrich17}
{Sch{\"o}nrich} R.,  {McMillan} P.~J.,  2017, \mn@doi [\mnras]
  {10.1093/mnras/stx093}, \href
  {https://ui.adsabs.harvard.edu/abs/2017MNRAS.467.1154S} {467, 1154}

\bibitem[\protect\citeauthoryear{{Sembach} et~al.,}{{Sembach}
  et~al.}{2003}]{sembach03}
{Sembach} K.~R.,  et~al., 2003, \mn@doi [\apjs] {10.1086/346231}, \href
  {http://adsabs.harvard.edu/abs/2003ApJS..146..165S} {146, 165}

\bibitem[\protect\citeauthoryear{{Shin}, {Stone}  \& {Snyder}}{{Shin}
  et~al.}{2008}]{shin08}
{Shin} M.-S.,  {Stone} J.~M.,   {Snyder} G.~F.,  2008, \mn@doi [\apj]
  {10.1086/587775}, \href
  {https://ui.adsabs.harvard.edu/abs/2008ApJ...680..336S} {680, 336}

\bibitem[\protect\citeauthoryear{{Shull}, {Jones}, {Danforth}  \&
  {Collins}}{{Shull} et~al.}{2009}]{shull09}
{Shull} J.~M.,  {Jones} J.~R.,  {Danforth} C.~W.,   {Collins} J.~A.,  2009,
  \mn@doi [\apj] {10.1088/0004-637X/699/1/754}, \href
  {https://ui.adsabs.harvard.edu/abs/2009ApJ...699..754S} {699, 754}

\bibitem[\protect\citeauthoryear{{Smith}}{{Smith}}{1963}]{smith63}
{Smith} G.~P.,  1963, \bain, \href
  {https://ui.adsabs.harvard.edu/abs/1963BAN....17..203S} {17, 203}

\bibitem[\protect\citeauthoryear{{Sormani} \& {Sobacchi}}{{Sormani} \&
  {Sobacchi}}{2019}]{sormani19}
{Sormani} M.~C.,  {Sobacchi} E.,  2019, \mn@doi [\mnras]
  {10.1093/mnras/stz793}, \href
  {https://ui.adsabs.harvard.edu/abs/2019MNRAS.486..215S} {486, 215}

\bibitem[\protect\citeauthoryear{{Sormani}, {Sobacchi}, {Pezzulli}, {Binney}
  \& {Klessen}}{{Sormani} et~al.}{2018}]{sormani18}
{Sormani} M.~C.,  {Sobacchi} E.,  {Pezzulli} G.,  {Binney} J.,   {Klessen}
  R.~S.,  2018, \mn@doi [\mnras] {10.1093/mnras/sty2500}, \href
  {https://ui.adsabs.harvard.edu/abs/2018MNRAS.481.3370S} {481, 3370}

\bibitem[\protect\citeauthoryear{{Sparre}, {Pfrommer}  \& {Ehlert}}{{Sparre}
  et~al.}{2020}]{sparre20}
{Sparre} M.,  {Pfrommer} C.,   {Ehlert} K.,  2020, \mn@doi [\mnras]
  {10.1093/mnras/staa3177}, \href
  {https://ui.adsabs.harvard.edu/abs/2020MNRAS.499.4261S} {499, 4261}

\bibitem[\protect\citeauthoryear{{Stanimirovi{\'c}}, {Dickey}, {Kr{\v{c}}o}  \&
  {Brooks}}{{Stanimirovi{\'c}} et~al.}{2002}]{stanimirovic02}
{Stanimirovi{\'c}} S.,  {Dickey} J.~M.,  {Kr{\v{c}}o} M.,   {Brooks} A.~M.,
  2002, \mn@doi [\apj] {10.1086/341892}, \href
  {https://ui.adsabs.harvard.edu/abs/2002ApJ...576..773S} {576, 773}

\bibitem[\protect\citeauthoryear{{Sun} \& {Reich}}{{Sun} \&
  {Reich}}{2010}]{sunreich10}
{Sun} X.-H.,  {Reich} W.,  2010, \mn@doi [Research in Astronomy and
  Astrophysics] {10.1088/1674-4527/10/12/009}, \href
  {http://adsabs.harvard.edu/abs/2010RAA....10.1287S} {10, 1287}

\bibitem[\protect\citeauthoryear{{Tepper-Garc{\'\i}a} \&
  {Bland-Hawthorn}}{{Tepper-Garc{\'\i}a} \&
  {Bland-Hawthorn}}{2018a}]{tepper-garcia18a}
{Tepper-Garc{\'\i}a} T.,  {Bland-Hawthorn} J.,  2018a, \mn@doi [\mnras]
  {10.1093/mnras/stx2680}, \href
  {https://ui.adsabs.harvard.edu/abs/2018MNRAS.473.5514T} {473, 5514}

\bibitem[\protect\citeauthoryear{{Tepper-Garc{\'\i}a} \&
  {Bland-Hawthorn}}{{Tepper-Garc{\'\i}a} \&
  {Bland-Hawthorn}}{2018b}]{tepper-garcia18b}
{Tepper-Garc{\'\i}a} T.,  {Bland-Hawthorn} J.,  2018b, \mn@doi [\mnras]
  {10.1093/mnras/sty1359}, \href
  {https://ui.adsabs.harvard.edu/abs/2018MNRAS.478.5263T} {478, 5263}

\bibitem[\protect\citeauthoryear{{Tepper-Garc{\'{\i}}a}, {Bland-Hawthorn}  \&
  {Sutherland}}{{Tepper-Garc{\'{\i}}a} et~al.}{2015}]{tepper-garcia15}
{Tepper-Garc{\'{\i}}a} T.,  {Bland-Hawthorn} J.,   {Sutherland} R.~S.,  2015,
  \mn@doi [\apj] {10.1088/0004-637X/813/2/94}, \href
  {http://adsabs.harvard.edu/abs/2015ApJ...813...94T} {813, 94}

\bibitem[\protect\citeauthoryear{{Teyssier}}{{Teyssier}}{2002}]{teyssier02}
{Teyssier} R.,  2002, \mn@doi [\aap] {10.1051/0004-6361:20011817}, \href
  {https://ui.adsabs.harvard.edu/abs/2002A&A...385..337T} {385, 337}

\bibitem[\protect\citeauthoryear{{Thom}, {Peek}, {Putman}, {Heiles}, {Peek}  \&
  {Wilhelm}}{{Thom} et~al.}{2008}]{thom08}
{Thom} C.,  {Peek} J.~E.~G.,  {Putman} M.~E.,  {Heiles} C.,  {Peek} K.~M.~G.,
  {Wilhelm} R.,  2008, \mn@doi [\apj] {10.1086/589960}, \href
  {http://adsabs.harvard.edu/abs/2008ApJ...684..364T} {684, 364}

\bibitem[\protect\citeauthoryear{{Unger} \& {Farrar}}{{Unger} \&
  {Farrar}}{2019}]{unger19}
{Unger} M.,  {Farrar} G.,  2019, in European Physical Journal Web of
  Conferences. p. 04005 (\mn@eprint {arXiv} {1901.04720}),
  \mn@doi{10.1051/epjconf/201921004005}

\bibitem[\protect\citeauthoryear{{Wakker}}{{Wakker}}{2001}]{wakker01}
{Wakker} B.~P.,  2001, \mn@doi [\apjs] {10.1086/321783}, \href
  {http://adsabs.harvard.edu/abs/2001ApJS..136..463W} {136, 463}

\bibitem[\protect\citeauthoryear{{Williams}, {de Geus}  \& {Blitz}}{{Williams}
  et~al.}{1994}]{williams94}
{Williams} J.~P.,  {de Geus} E.~J.,   {Blitz} L.,  1994, \mn@doi [\apj]
  {10.1086/174279}, \href
  {https://ui.adsabs.harvard.edu/abs/1994ApJ...428..693W} {428, 693}

\bibitem[\protect\citeauthoryear{{Zhang}, {Thompson}, {Quataert}  \&
  {Murray}}{{Zhang} et~al.}{2017}]{zhang17}
{Zhang} D.,  {Thompson} T.~A.,  {Quataert} E.,   {Murray} N.,  2017, \mn@doi
  [\mnras] {10.1093/mnras/stx822}, \href
  {https://ui.adsabs.harvard.edu/abs/2017MNRAS.468.4801Z} {468, 4801}

\bibitem[\protect\citeauthoryear{{Zheng}, {Peek}, {Werk}  \& {Putman}}{{Zheng}
  et~al.}{2017}]{zheng17}
{Zheng} Y.,  {Peek} J.~E.~G.,  {Werk} J.~K.,   {Putman} M.~E.,  2017, \mn@doi
  [\apj] {10.3847/1538-4357/834/2/179}, \href
  {https://ui.adsabs.harvard.edu/abs/2017ApJ...834..179Z} {834, 179}

\bibitem[\protect\citeauthoryear{{van de Voort}, {Bieri}, {Pakmor},
  {G{\'o}mez}, {Grand}  \& {Marinacci}}{{van de Voort}
  et~al.}{2021}]{vandevoort21}
{van de Voort} F.,  {Bieri} R.,  {Pakmor} R.,  {G{\'o}mez} F.~A.,  {Grand} R.
  J.~J.,   {Marinacci} F.,  2021, \mn@doi [\mnras] {10.1093/mnras/staa3938},
  \href {https://ui.adsabs.harvard.edu/abs/2021MNRAS.501.4888V} {501, 4888}

\bibitem[\protect\citeauthoryear{{van den Bergh}}{{van den
  Bergh}}{1962}]{vandenbergh62}
{van den Bergh} S.,  1962, \mn@doi [\aj] {10.1086/108757}, \href
  {http://adsabs.harvard.edu/abs/1962AJ.....67..486V} {67, 486}

\makeatother
\end{thebibliography}

\appendix
\section{Comoving cloud reference frame}
\label{sec:refframe}
As mentioned previously, we run our simulations in a reference frame that is approximately comoving with the cloud material. Specifically, we subtract the $z$-velocity $\langle v_z \rangle$ of the centre of mass of the cloud material from the $z$-velocity in every cell in the simulation after every fourth time step on the coarsest AMR level. We keep track of the cloud material through a passive scalar $C$ (see Section \ref{sec:nummethod}). Hence, $\langle v_z \rangle$ (as discretised on the mesh so that integrals become sums) is given by
\begin{equation}
\langle v_z \rangle = \frac{\sum_i \rho_i C_i v_{z,i} V_i}{\sum_i \rho_i C_i V_i},
\end{equation}
where $V$ is volume and the subscript $i$ refers to the quantity in the $i$th cell. This generally keeps the cloud's centre of mass velocity below $1 \kms$ in the simulation. We refer to this reference frame as the `cloud frame' and the frame where no velocity subtraction is applied as the `corona frame'. While the cloud material, as tracked by $C$, is initially equivalent to the cold gas, this changes during the simulation as condensation creates additional cold gas and, conversely, as some initially cold gas is adiabatically heated. However, as cooling dominates over heating and is driven by mixing, the passive scalar generally tracks the cold gas throughout the simulations \citep[see also Figure 5 of][]{gronnow18}. Similar frame changing methods have been previously employed in a number of wind tunnel and shock-cloud simulations in the literature \citep[e.g.][]{shin08,mccourt15,scannapieco15,dutta19}. This is often used to follow the cloud without requiring the simulation domain to cover the entire length of its trajectory. In our case, our domain is not much smaller than the length of the entire trajectory because we need to also keep the wake of the cloud within the domain for our purposes. Our use of AMR also means that a large domain is not a computational issue. Instead, our primary motivation for using frame changing is that not including the denser inner parts of the corona during most of the simulations leads to better numerical stability of the corona. In addition, \cite{mccourt15} found that the cloud frame leads to less numerical diffusion errors from the advection of the cloud across the mesh. This is due to the overall lower velocity difference between the mesh and the cloud material in the cloud frame.

\begin{figure}
    \centering
    \includegraphics[width=0.49\textwidth]{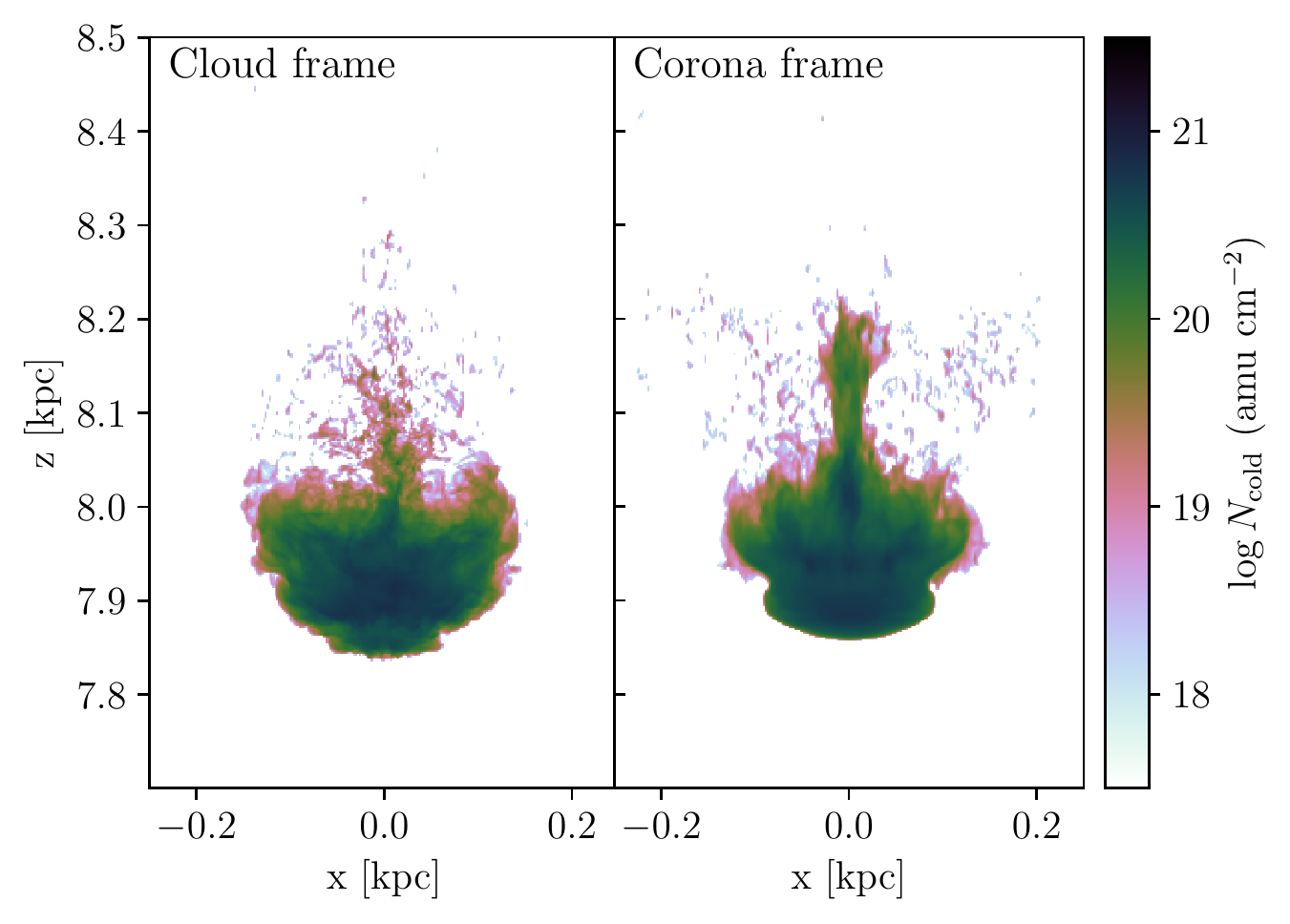}
    \caption{Projection along $y$ of the logarithmic density of cold gas ($T<2\times 10^4$ K) in simulation HD at $t=35$ Myr for a simulation in the cloud frame (left) and in the corona frame (right).}
    \label{fig:refframesdens}
\end{figure}

To assess the reliability of this method for our simulations, we compare the cold gas mass in simulations MHD-S and HD in the cloud frame to the corona frame where we do not subtract the cloud's velocity. For the corona frame we use a domain of $16$ kpc along each direction with heights spanning 1 kpc $\leq z \leq 17$ kpc with the AMR levels adjusted accordingly such that each level has the same resolution as in the cloud frame. Due to the instability mentioned previously, we could only run MHD-S in the corona frame until $t\approx 35$ Myr.

\begin{figure}
    \centering
    \includegraphics[width=0.49\textwidth]{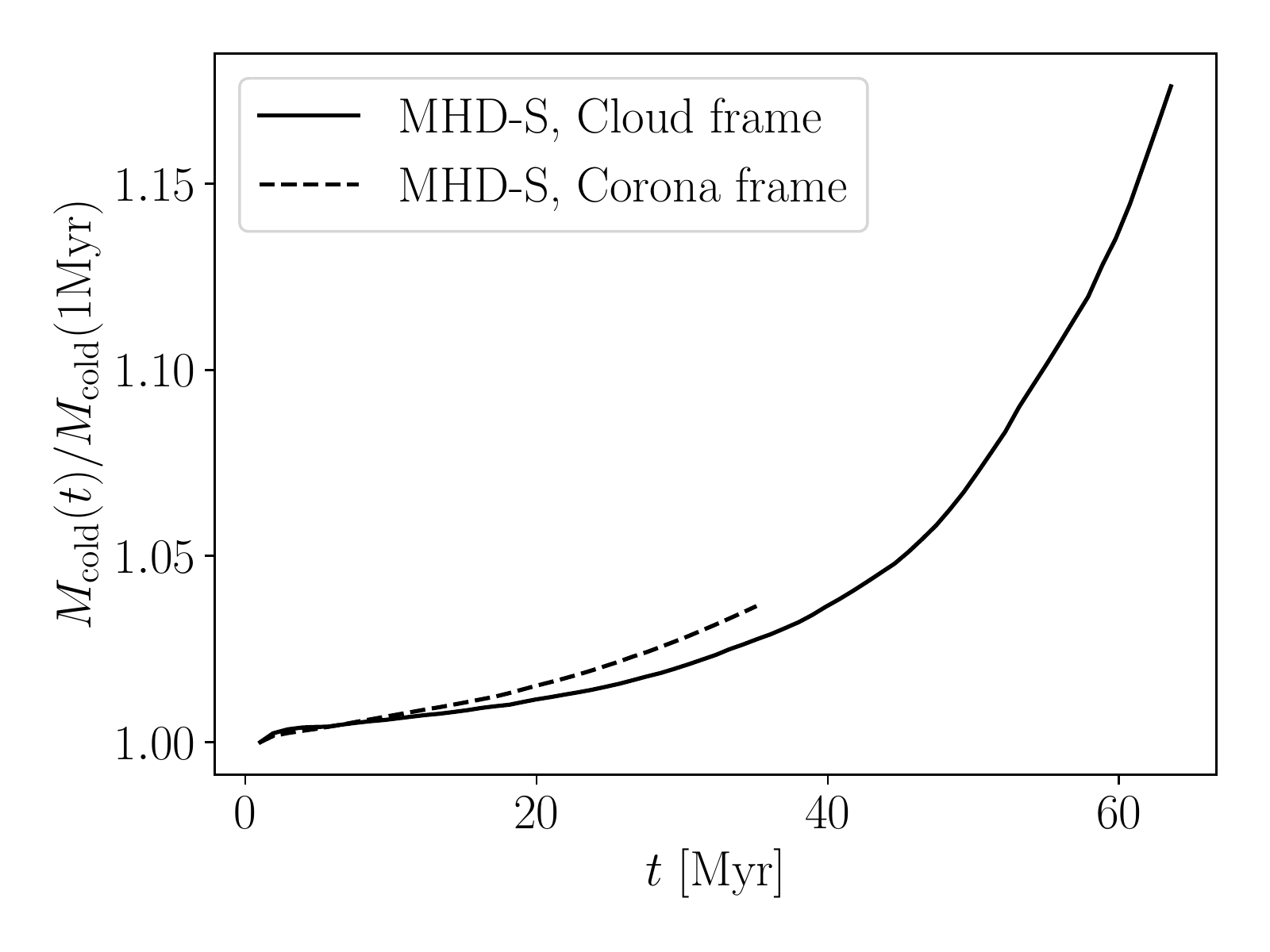}
    \includegraphics[width=0.49\textwidth]{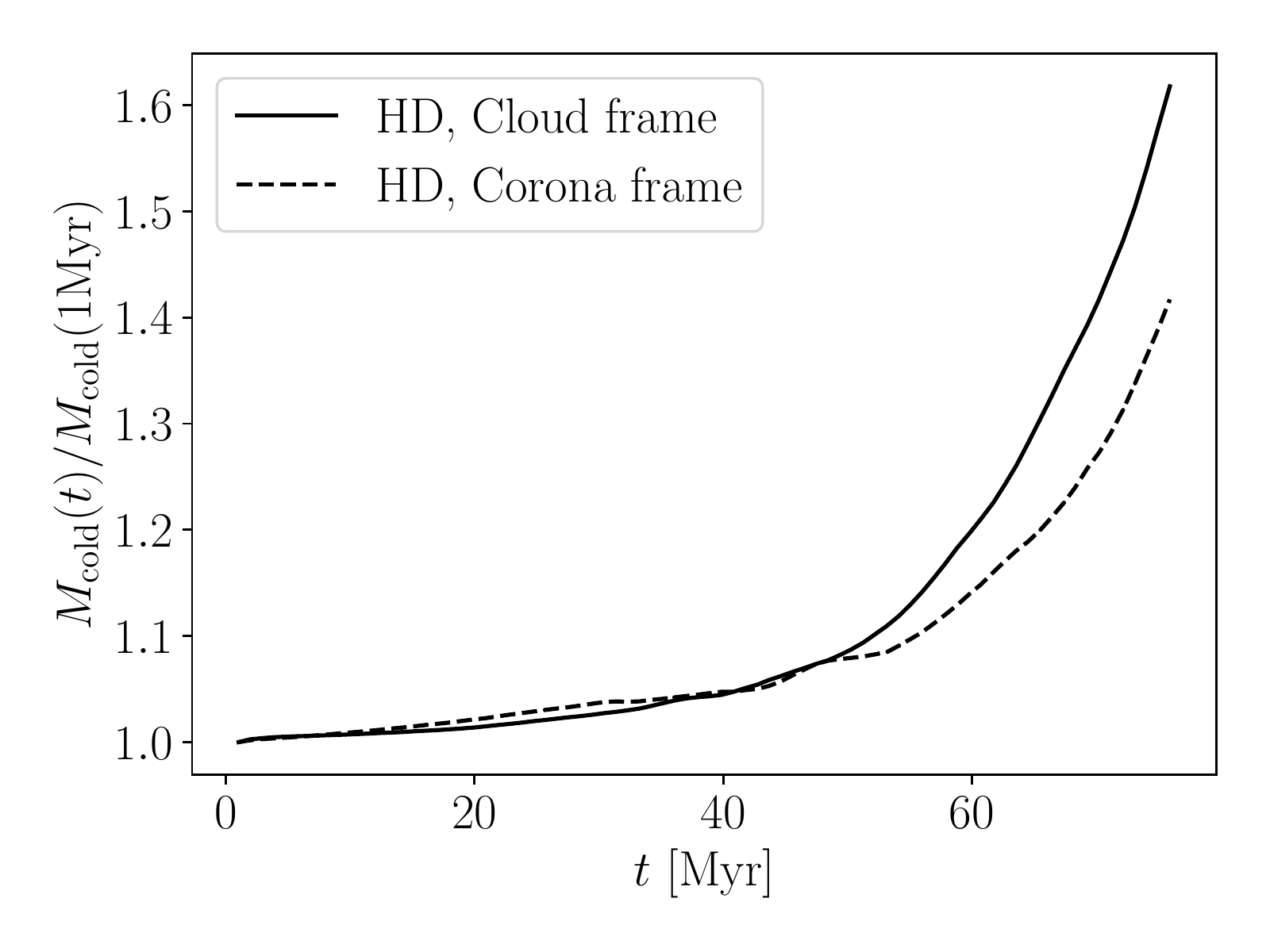}
    \caption{Cold gas mass evolution for the simulation with a strong magnetic field (top) and without a magnetic field (bottom) in the cloud frame (solid lines) and corona frame (dashed lines).}
    \label{fig:refframesmcold}
\end{figure}

In Figure \ref{fig:refframesdens} we show a projection of the cold gas density in simulation HD at $t=35$ Myr. The cloud looks overall similar but more smoothed out in the corona frame, as expected from the additional numerical diffusion. We show the cold gas mass evolution in the two reference frames in simulation MHD-S and HD in the top and bottom panels of Figure \ref{fig:refframesmcold}, respectively. As can be seen, the evolution of the cold gas mass in the two frames are qualitatively similar. In particular, it increases with time apart from a few insignificant drops in simulation HD in the corona frame. For simulation HD the cold gas mass is slightly higher in the corona frame than in the cloud frame at early times. Hence, the effect of the additional numerical diffusion is to cause more cooling due to the smoothing of temperature gradients at early times. However, the mass of cold gas becomes slightly lower at later times presumably because the smoothing of the cloud-corona interface reduces the growth of KH and RT instability that drives the stripping and mixing. This is similar to the effect seen when physical diffusion is added by including thermal conduction \citep[see e.g.][]{armillotta17,kooij21}. The fact that the lower resolution simulation MHD-Sl likewise overestimates(underestimates) the mass of cold gas at early(late) times is also consistent with this interpretation (see Section \ref{sec:resolution}).

The centre of mass velocity being a mass-weighted average is dominated by the main cloud remnants. While this ensures that most of the mass is advecting slowly across the mesh, the clumps in the wake, which are almost comoving with the corona in simulation MHD-S, have a high velocity in the cloud frame. However, these clumps should not be significantly stripped due to their low velocity w.r.t. the surrounding gas and are not resolved in any case. Hence, the additional numerical diffusion of these clumps in the cloud frame compared to the corona frame should not be a problem so long as it is not severe enough to disperse them. This is clearly not the case as a large population of these cloudlets exists down to our minimum mass and size thresholds. Simulation MHD-S in the corona frame ends too early for a significant mass of low velocity (in the corona frame) condensed gas as seen at $t=64$ Myr in Figure \ref{fig:coldgasvel} to have formed. However, we do see a small bump at low velocities in the velocity distribution of the cold gas mass similar to that of the cloud frame at $t=35$ Myr.

In summary, we find that the approximately comoving cloud frame, that we use in our standard simulations, largely agrees with the corona frame but appears to have the advantage of additional stability and lower numerical diffusion.

\label{lastpage}
\end{document}